\documentclass[aps,prb,reprint,groupedaddress]{revtex4-1}
\usepackage{graphicx}
\usepackage{hyperref}
\usepackage{amsfonts}
\usepackage{amsmath,bm}
\usepackage{amssymb}
\usepackage{graphicx}
\usepackage{subfigure}
\usepackage{longtable}
\usepackage{color}

\draft
\begin{document}


\title{Entropy variation rate divided by temperature always decreases}



\author{T.M. Shih$^{1,2}$}
\email[]{tmshih@xmu.edu.cn}
\author{Z.J. Gao$^{1}$}
\author{H. Merlitz$^{1,3}$}
\author{L. Rondoni$^{4,5}$}
\author{P. J. Pagni$^{6}$}
\author{Z. Chen$^{7}$}
\email[]{chenz@xmu.edu.cn}
\affiliation{$1$.Department of Physics, Xiamen University, Xiamen, China 361005}
\affiliation{$2$.Institute for Complex Adaptive Matter, University of California, Davis, CA 95616, USA}
\affiliation{$3$.Leibniz Institute for Polymer Research, Dresden, Germany}
\affiliation{$4$.Dipartimento di Scienze Matematiche and Graphene@Polito Lab Politecnico di Torino, Corso Duca degli Abruzzi 24, 10129 Torino, Italy}
\affiliation{$5$.INFN, Sezione di Torino, Via P. Giuria 1, 10125 Torino, Italy}
\affiliation{$6$.Department of Mechanical Engineering, University of California, Berkeley, CA 94720, USA}
\affiliation{$7$.Department of Electronic Science, Fujian Engineering Research Center for Solid-state Lighting,
State Key Laboratory for Physical Chemistry of Solid Surfaces, Xiamen University, Xiamen,
China 361005}


\date{\today}

\begin{abstract}

\end{abstract}

\pacs{}

\maketitle


For an isolated assembly that comprises a system and its surrounding reservoirs, the total entropy ($S_{a}$) always monotonically increases as time elapses\cite{1, 2, 3}. This phenomenon is known as the second law of thermodynamics ($S_{a}\geq0$). Here we analytically prove that, unlike the entropy itself, the entropy variation rate ($B=dS_{a}/dt$) defies the monotonicity for multiple reservoirs ($n\geq2$). In other words, there always exist minima. For example, when a system is heated by two reservoirs from $T=300\,K$ initially to $T=400\,K$ at the final steady state, $B$ decreases steadily first. Then suddenly it turns around and starts to increases at $387\,K$ until it reaches its steady-state value, exhibiting peculiar dipping behaviors. In addition, the crux of our work is the proof that a newly-defined variable, $B/T$, always decreases. Our proof involves the Newton¡¯s law of cooling, in which the heat transfer coefficient is assumed to be constant. These theoretical macro-scale findings are validated by numerical experiments using the Crank-Nicholson method, and are illustrated with practical examples. They constitute an alternative to the traditional second-law statement, and may provide useful references for the future micro-scale entropy-related research.\par

\begin{figure}[h]
\subfigure{\includegraphics[width=0.23\textwidth]{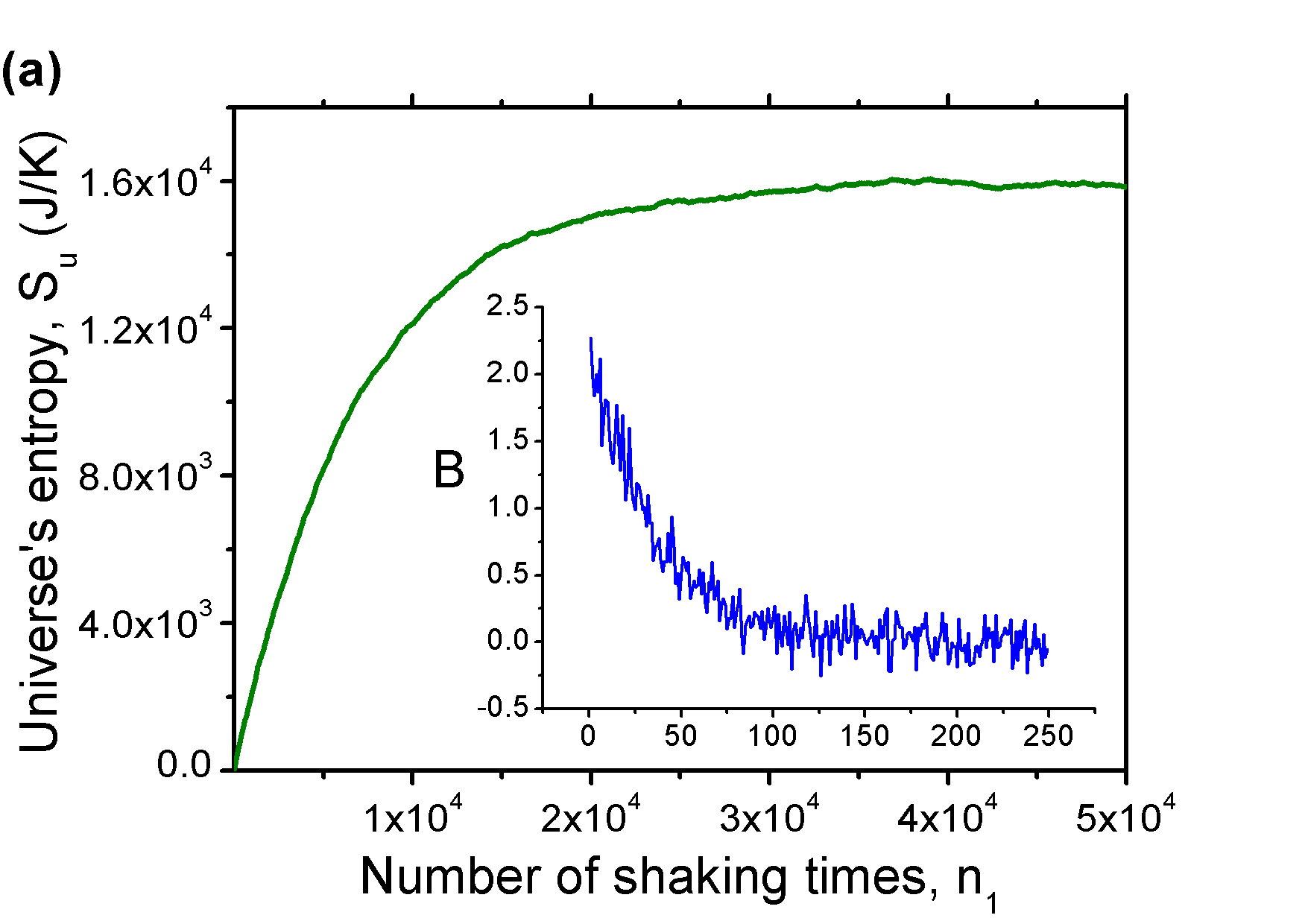}}
\subfigure{\includegraphics[width=0.23\textwidth]{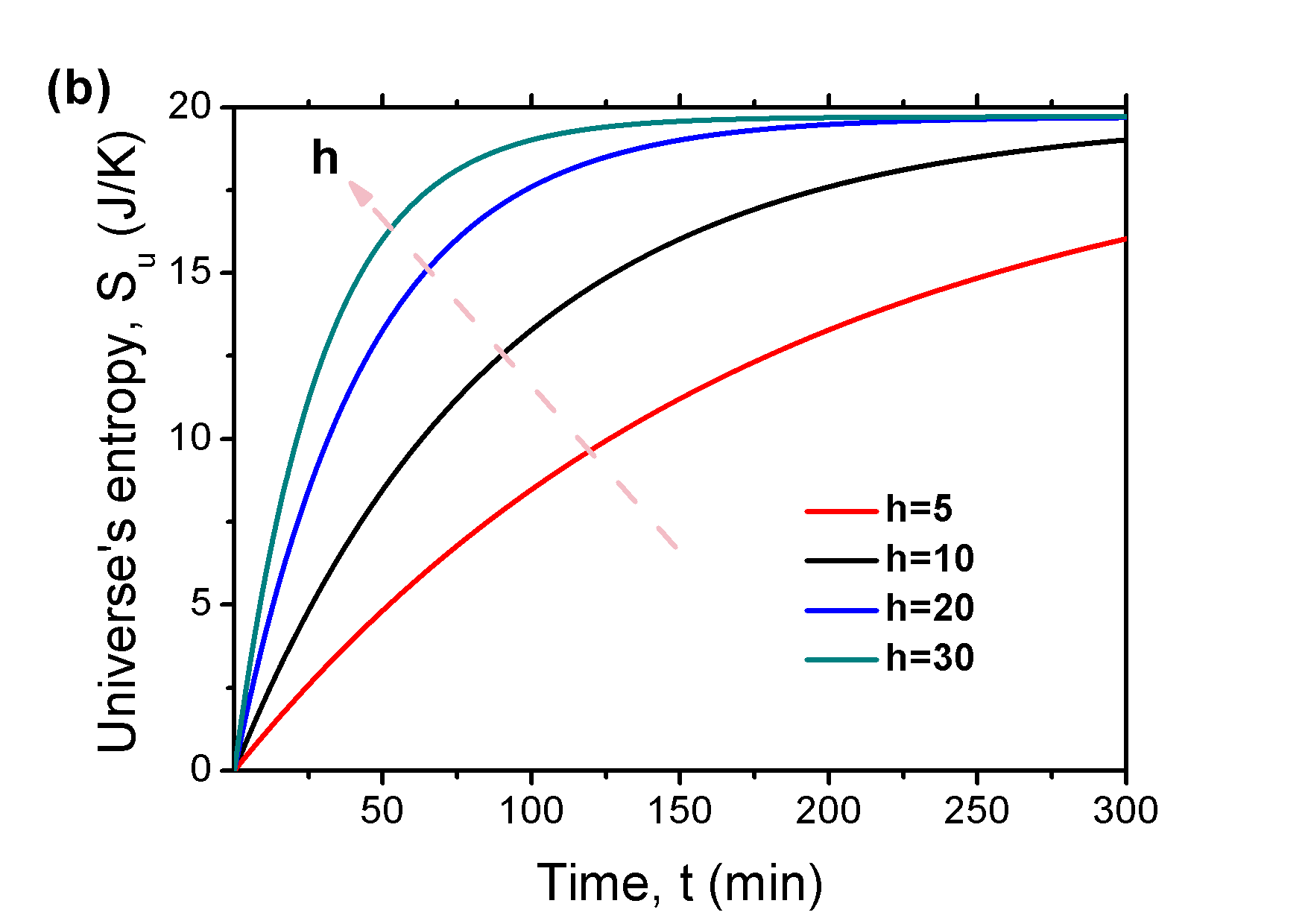}}
\caption{\textbf{Trends for the entropy $\bm S_{\bm u}$ and its variation rate, $\bm B\bm (\bm t\bm )$, in both microscopic and macroscopic views. $\bm a$,} $S_{u}$ and $B$ versus number of shakings. {$\bm b$,} entropy of universe (bottle of beer $+$ living room) versus time, parametrized in the heat-transfer coefficient, $h\,(W/m^{2}K)$, with $m= 2\,kg,\,c_{v}=4180\,J/kgK, \text{and} A=0.0768\,m^{2}.$}
\label{PIC1}
\end{figure}

When a car decelerates on the road, the distance traveled by the car continues to increases monotonically, while the car speed decreases monotonically. Likewise, consider a dimpled tray, on which $N/2$ ($N\gg1$) black marbles and $N/2$ white marbles are orderly placed. When shaking the tray, we observe from afar that its grayness (Supplementary S-$1$) gradually increases, while its variation rate diminishes asymptotically to zero (Fig. \ref{PIC1}$a$). In macro-scales, the entropy of an assembly, $S_{a}$, consisting of a system and a single reservoir, always increases or remains constant. Its variation rate, $B=dS_{a}/dt$, always decreases and diminishes to zero eventually. For example,
a bottle of beer at $5^{\circ}C$ taken out from the refrigerator, is placed in the living room at $25^{\circ}C$ (assembly $=$ bottle of beer $+$ living room). The warming rate of the beer, $J$ (watts), is governed by
\begin{equation}\label{1}
J(t)=Ah(T_{R}-T(t)),
\end{equation}
known as Newton's law of cooling\cite{4,5} (or warming), where $A$ is the surface area of the bottle; $h$ the heat transfer coefficient assumed to be constant; $T_{R}$ the temperature of the living-room air (reservoir); and $T(t)$ is the temperature of the beer, which is approximately uniform due to convective mixing inside the bottle. Based on first and second laws of thermodynamics\cite{6}, we obtain
\begin{equation}\label{2}
\frac{dU_{a}(t)}{dt}=J(t)
\end{equation}
and
\begin{equation}\label{3}
B(t)=J(t)\left(\frac{1}{T(t)}-\frac{1}{T_{R}}\right),
\end{equation}
where $U_{a}$ is the internal energy of the system (bottle of beer). Representative results within a period of $10\,h$ are listed in Table \ref{table1}, showing that $S_{a}$, or microscopic disorder\cite{7,8}, increases steadily as time elapses. As the process is assumed to occur closely to equilibrium, the thermodynamic entropy equals the non-equilibrium statistical mechanical entropies (Boltzmann entropy and Lebowitz entropy among others, depending on the problem\cite{9,10,11}). Furthermore, during the first hour, $S_{a}$ has increased by $9.67\,J/K$. The rate $B(t)$ shown on the rightmost column decreases monotonically, and diminishes to zero after $10\,h$. If we bring a fan to blow on the bottle, resulting in an increase of $h$ from $10\,W/m^{2}K$ to $20$, $S_{a}$ will increase by $14.58\,J/K$ (Fig. \ref{PIC1}$b$). Likewise, it is reasonable to imagine and believe that another assembly consisting of a house and the atmosphere under the sudden attack of a hurricane\cite{12} will experience a higher entropy increase than in a calm day. These trends pose no surprises.\par

\begin{figure}
\subfigure{\includegraphics[width=0.23\textwidth]{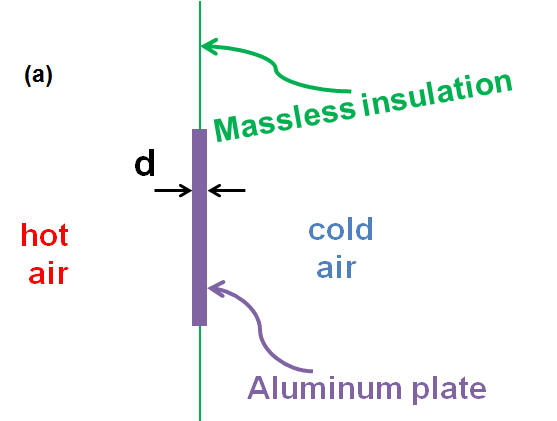}}
\subfigure{\includegraphics[width=0.23\textwidth]{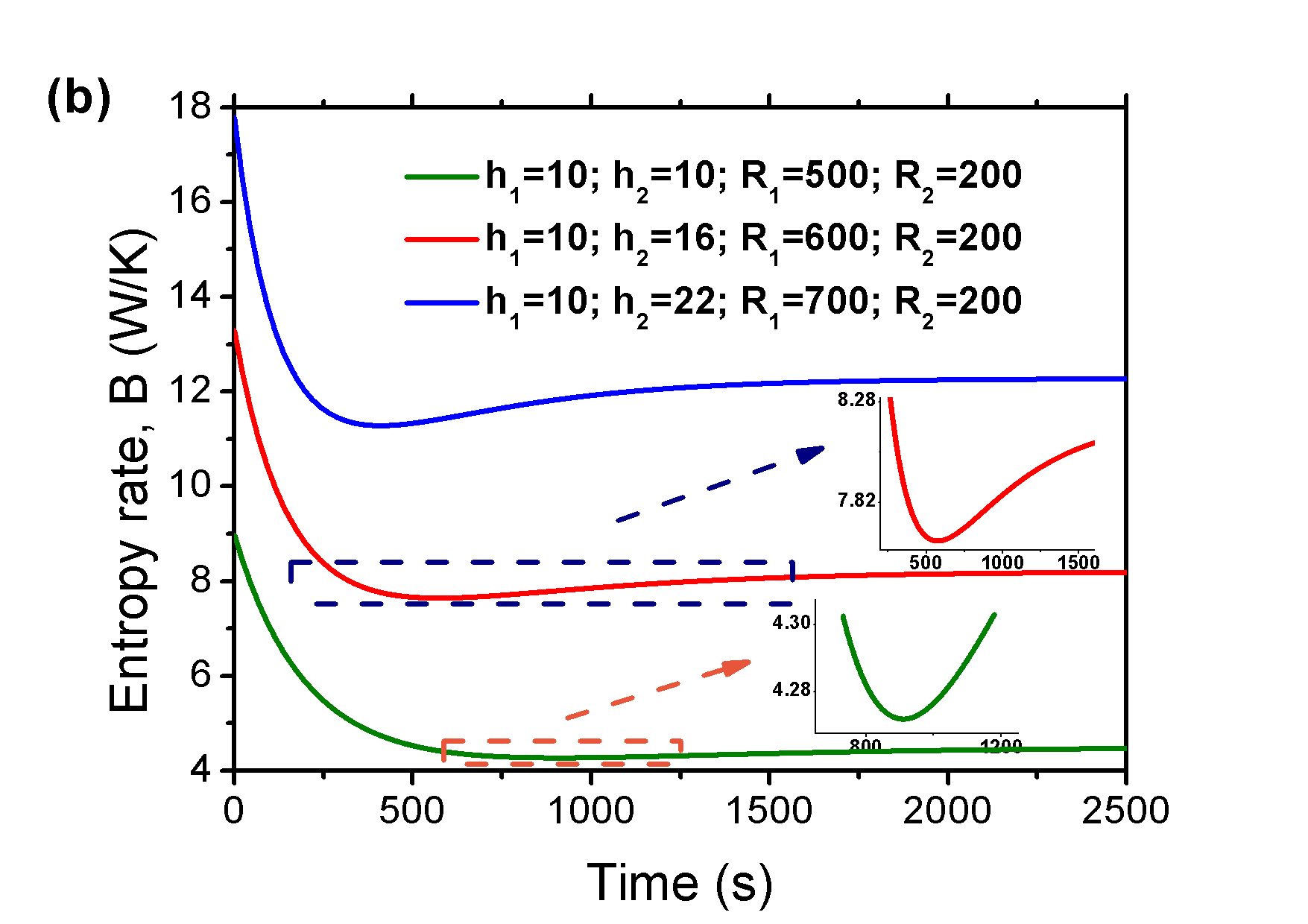}}\\
\subfigure{\includegraphics[width=0.23\textwidth]{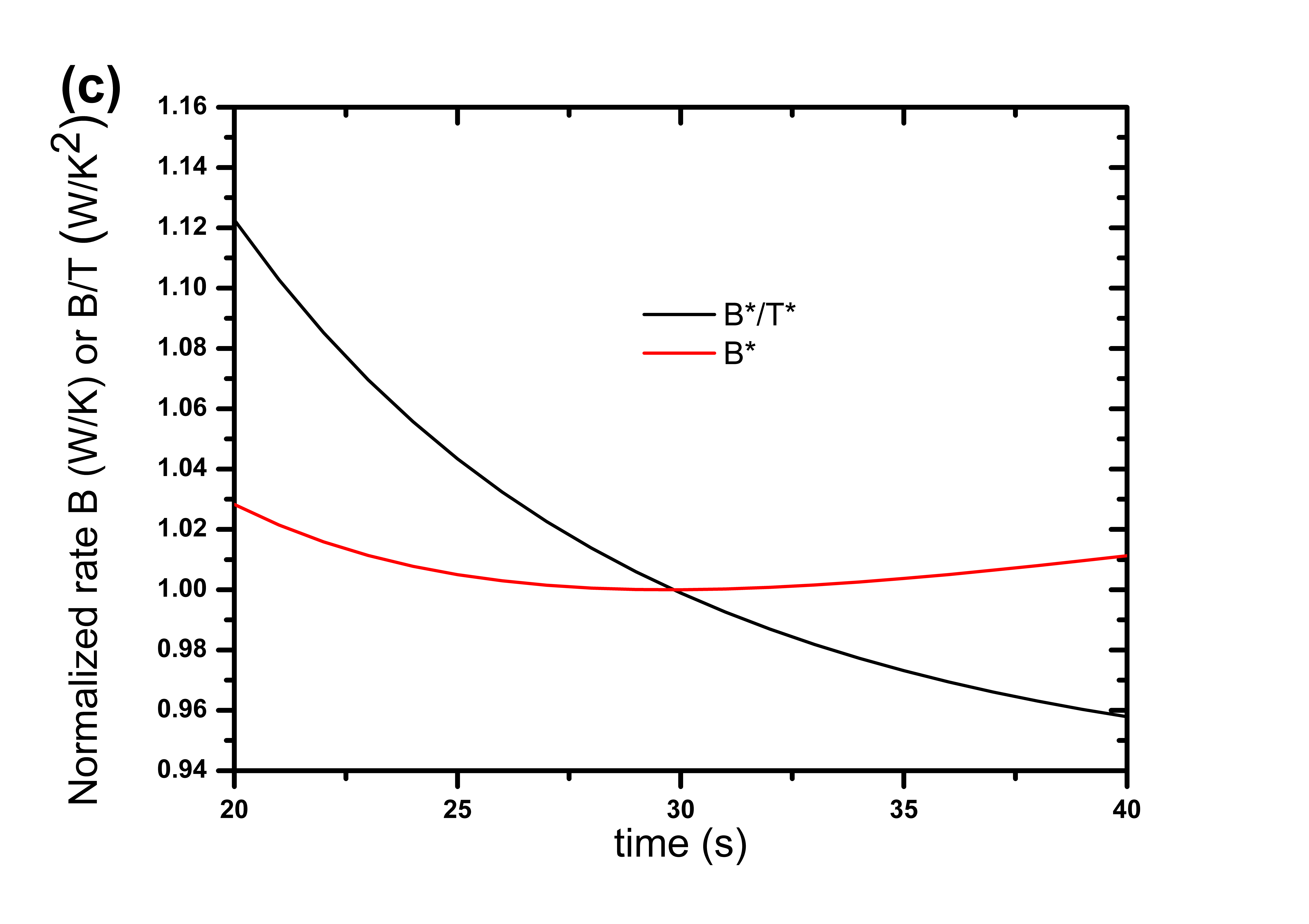}}
\subfigure{\includegraphics[width=0.23\textwidth]{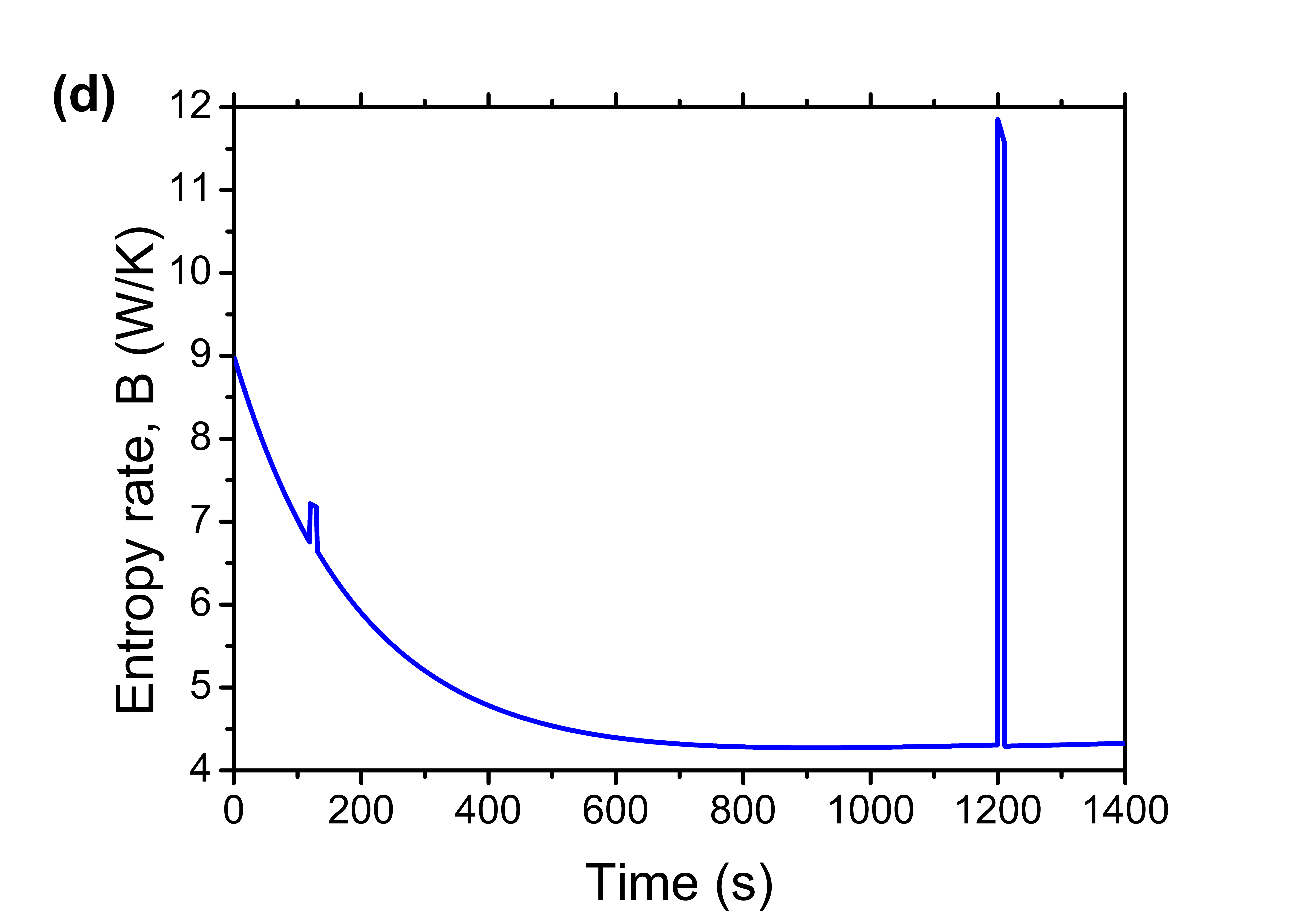}}
\subfigure{\includegraphics[width=0.23\textwidth]{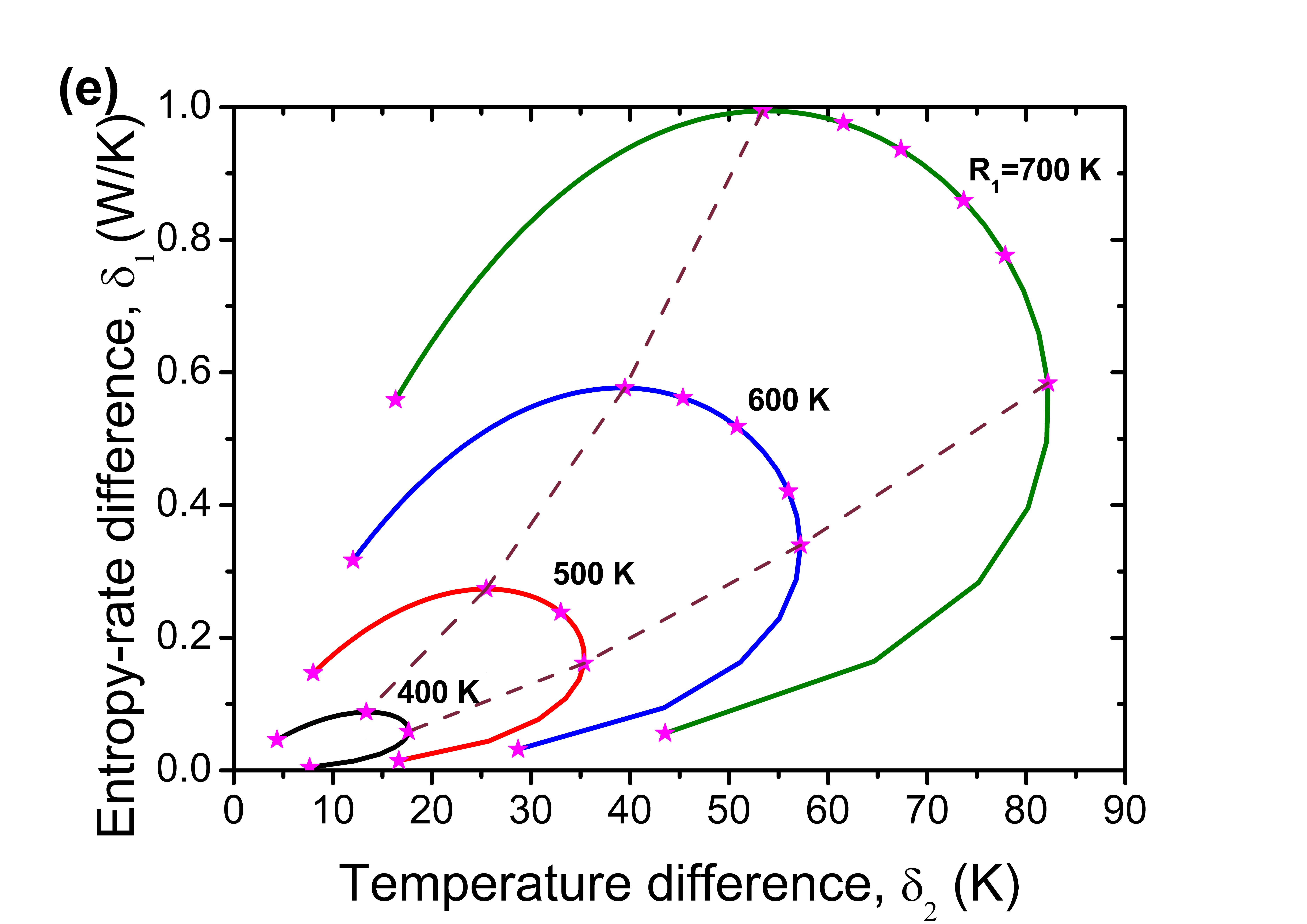}}
\subfigure{\includegraphics[width=0.23\textwidth]{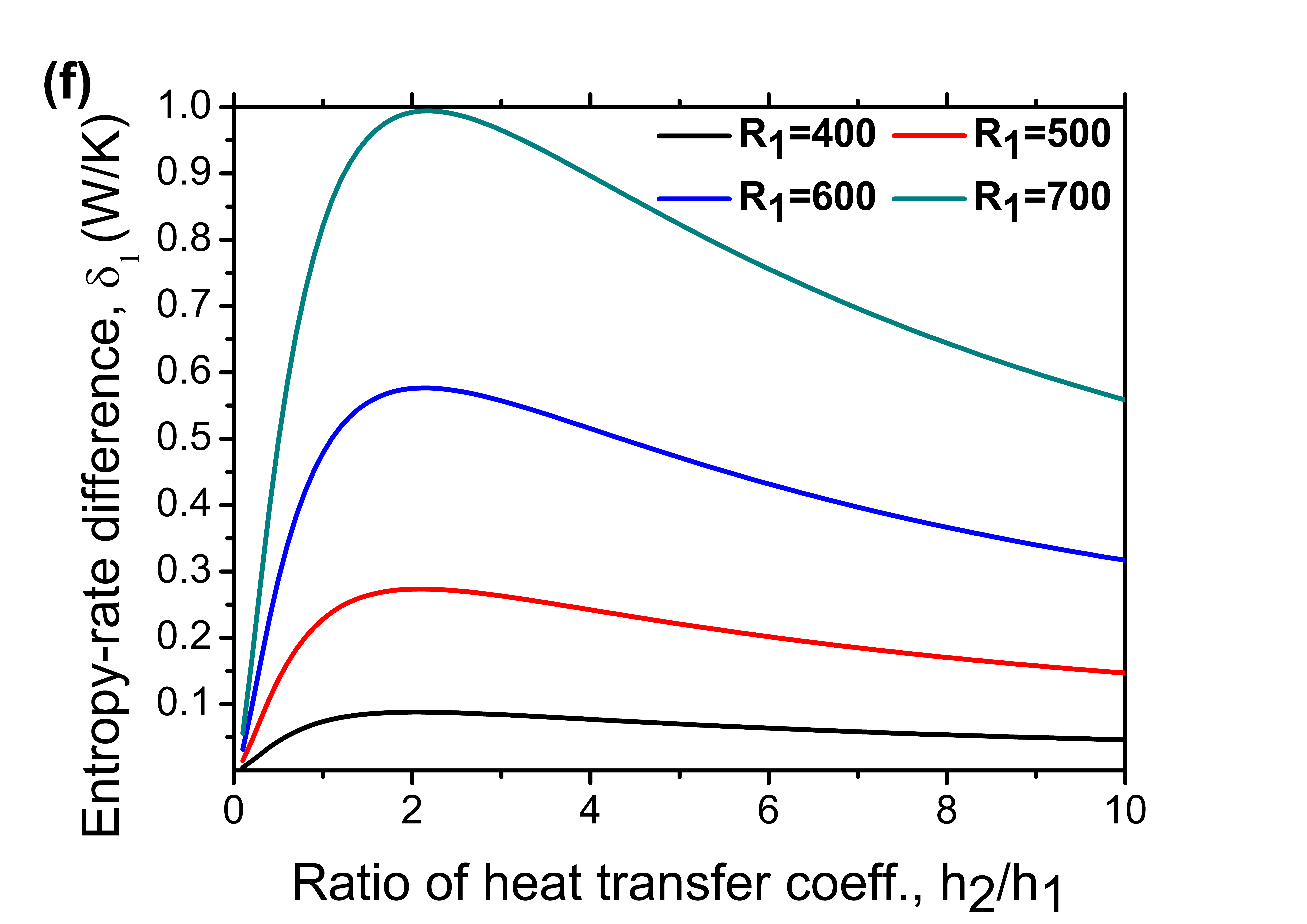}}
\subfigure{\includegraphics[width=0.23\textwidth]{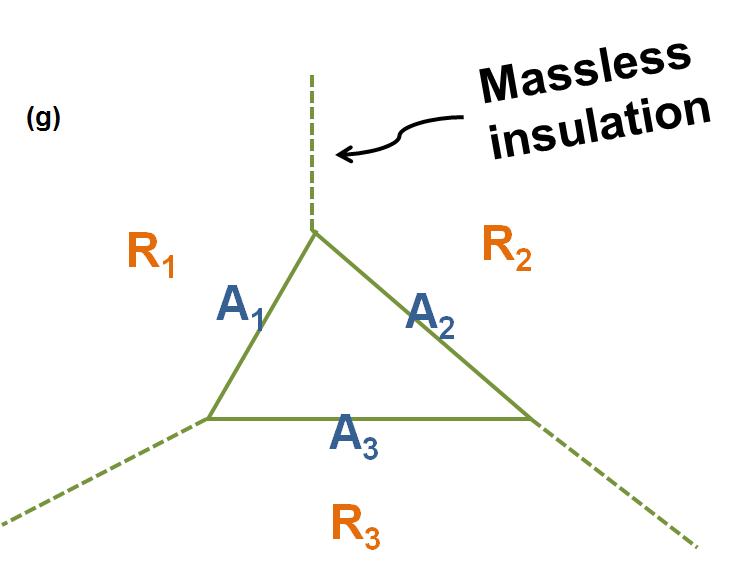}}
\subfigure{\includegraphics[width=0.23\textwidth]{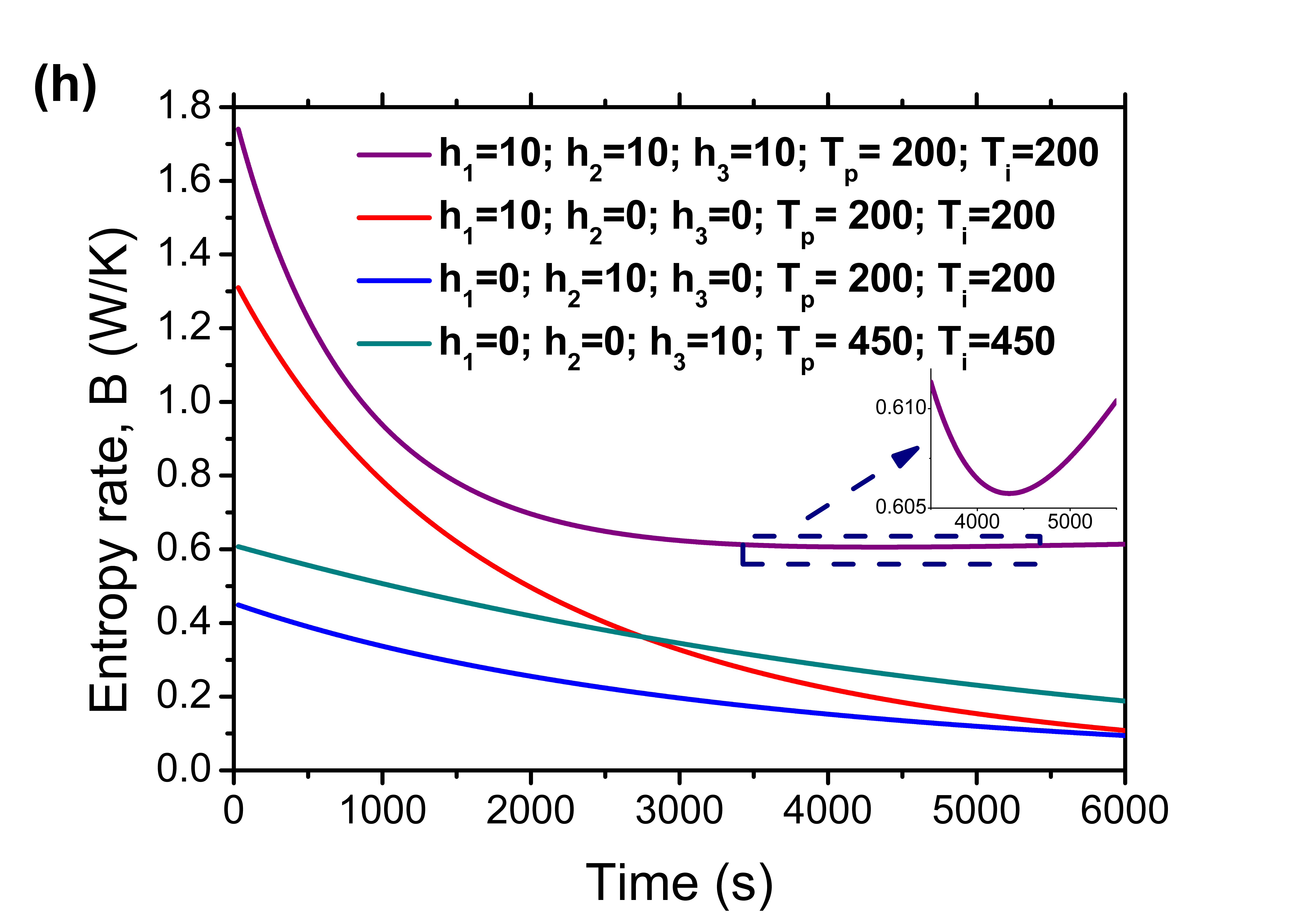}}
\caption{\textbf{System schematics and entropy-related results for universes of one system immersed in two or three reservoirs. $\bm a$,} A vertical aluminum plate exposed to two reservoirs. The cold reservoir is maintained at $R_{2}=200\,K$. The heat-transfer coefficient on the hot side is maintained at $h_{1}=10\,W/m^{2}K$. The temperature of the hot reservoir ranges from $R_{1}=300\,K$ to $700\,K$. Initially, the temperature of the plate is at $200\,K$. $\bm b,$ Rate $B(t)$ versus time for various values of $R_{1}$ and $h_{1}$. For $h_{2}=10$ and $R_{1}=500\,K$, the existence of $B_{min}$ is inconspicuous, requiring an inset figure to be revealed. $\bm c,$ The spike in the entropy accelerating regime and the tiny burst in the entropy decelerating regime are both induced by an external perturbation (a sudden increase of $h_{2}$ from $10$ to $40$ for a period of $10\,s$). $\bm d,$ Difference in $B_{ss}$ and $B_{min}$ versus difference in $T_{ss}$ and $T_{if}$, parametrized in $R_{1}$, which varies from $400\,K$ to $700\,K$. For all curves, $h_{2}/h_{1}$ varies from $0.1$ to $10$ (at low ends of curves to high ends, not shown to avoid confusion). $\bm e,$ Difference in $B_{ss}$ and $B_{min}$ versus the ratio, $h_{2}/h_{1}$. $\bm f,$ System schematic of a triangular block exposed to three reservoirs, with $A_{1}=0.10\,m^{2}$, $A_{2}=0.0653\,m^{2}$, $A_{3}=0.0879\,m^{2}$, and $m=7.6366\,kg$. $\bm g,$ Entropy variation rate, $B$, versus time for various conditions.}
\label{PIC2}
\end{figure}

In microscales, the presence of two reservoirs has been investigated for electron transfers between different temperatures and/or electrochemical potentials\cite{13}, along with a memory register via Landauer's principle\cite{14}, interacting with a single-electron box\cite{15}, and coupled with electrical charges\cite{16}. Macroscopically, it can be shown that, indeed, the existence of multiple reservoirs does offer interesting physical insights that are absent in cases of single reservoirs.
\begin{table}
\scriptsize
\begin{center}
\begin{tabular}{cccccc}
\hline
$t$             &$T_{beer}$ &$S_{beer}$ &$S_{R}$ &$S_{u}$ &$B$\\
             &$(K)$ &$(J/kg)$ &$(J/kg)$ &$(J/kg)$ &$(W/kg)$\\\hline
$0\,s$             &$278$           &$0$               &$0$           &$0$           &$n/a$\\
$1\,s$             &$278.002$      &$0.055$                &$-0.052$            &$3.706\times10^{-3}$            &$3.7060\times10^{-3}$\\
$2\,s$             &$278.004$              &$0.110$                &$-0.103$            &$7.411\times10^{-3}$            &$3.7053\times10^{-3}$\\\hline
$10\,s$             &$278.018$             &$0.552$                &$-0.515$            &$3.703\times10^{-2}$            &$3.6997\times10^{-3}$\\
$11\,s$             &$278.020$             &$0.607$                &$-0.566$            &$4.073\times10^{-2}$            &$3.6990\times10^{-3}$\\\hline
$1\,h$             &$283.630$          &$167.600$                &$-157.929$             &$9.671$          &$1.8757\times10^{-3}$\\
$10\,h$             &$297.266$          &$560.185$               &$-540.495$            &$19.690$           &$4.7000\times10^{-6}$\\\hline
\end{tabular}
\caption{\textbf{Selective values in time evolution for the mini-universe consisting of a bottle of beer and the living room.} The monotonicity of the entropy variation rate, $B$, is observed.}
\label{table1}
\end{center}
\end{table}
An aluminum plate is vertically placed, with its left and right faces immersed in reservoirs $1$ and $2$, respectively (Fig. \ref{PIC2}$a$). The value of Biot number, $hd/k$, for the plate is $2.1\times10^{-4}$, which is lower than the conventional criterion ($\approx0.001$), so that the heat conduction phenomenon inside the plate qualifies as a lumped-capacitance model (Supplementary S-$2$). The rate $B$ and the entropy acceleration can be derived (Supplementary S-$3$) as
\begin{equation}\label{4}
B=A_{1}h_{1}\left(\frac{R_{1}}{T}+\frac{T}{R_{1}}-2\right)+A_{2}h_{2}\left(\frac{R_{2}}{T}+\frac{T}{R_{2}}-2\right),
\end{equation}
and
\begin{equation}\label{5}
\begin{split}
\frac{d^{2}S_{a}}{dt^{2}}=&\left(\frac{dT}{dt}\right)\left(\frac{A_{2}h_{2}}{T^{2}R_{2}}\right)\left(T^{2}-R_{2}^{2}\right)\\
&-\left(\frac{dT}{dt}\right)\left(\frac{A_{1}h_{1}}{T^{2}R_{1}}\right)\left(R_{1}^{2}-T^{2}\right),
\end{split}
\end{equation}
where $R_{1}$ and $R_{2}$ denote temperatures of hot and cold reservoirs. The rate $B$ gradually levels off and diminishes, with nothing suspicious at the first glance (the lowest curve in Fig. \ref{PIC2}$b$). When enlarging the region near $t=975\,s$, however, we observe that actually a $B_{min}$ exists, implying that an inflection point, $S_{a}''=0$, also exists, as shown in the inset figure. This phenomenon, which does not happen in the presence of one single reservoir, calls for a physical interpretation. When subject to only one single (e.g., hot) reservoir, the system will gradually be warmed up, and eventually reach an equilibrium state without further irreversible entropy production\cite{17}. In Eq. (\ref{5}), $h_{2}$ can be set to zero. Since $dT/dt$ is always positive, and $R_{1}$ is always higher than $T$, $d^{2}S_{a}/dt^{2}$ always remains negative, indicating that the entropy variation rate should monotonically decrease and asymptotically diminish to zero. When two reservoirs co-exist, the plate simultaneously experiences two thermal influences, expressed by two terms in the right-hand side of Eq. (\ref{5}). Depending on the process being warming or cooling, $dT/dt$ should be either positive or negative. Regardless, $T$ is always higher than $R_{2}$, and $R_{1}$ is always higher than $T$, leading to the possibility that these two competing strengths become equal. The occurrence of this possibility dictates
\begin{equation}\label{6}
d^{2}S_{u}/dt^{2}=0,
\end{equation}
when
\begin{equation}\label{7}
T_{if}=\left(\frac{A_{1}h_{1}R_{1}+A_{2}h_{2}R_{2}}{A_{1}h_{1}/R_{1}+A_{2}h_{2}/R_{2}}\right)^{1/2},
\end{equation}
where the subscript ``if'' stands for ``inflection''. Equations (\ref{6}) and (\ref{7}) define the entropy inflection point for a given assembly that consists of one system and two reservoirs. In fact, the existence of the entropy inflection point is not merely a possibility; it is a certainty as $T$ reaches $T_{if}$ if the system is given with sufficient time to warm up. We can further prove (Supplementary S-$4$) that
\begin{equation}\label{8}
T_{if}\leq T_{ss}\text{ always for }n\geq 2,
\end{equation}
where $n$ is the number of reservoirs. The equality prevails only when $R_{1}=R_{2}$. If the initial temperature of the system is higher than $T_{if}$, the system will approach the steady state without passing through the inflection point. However, The existence of the inflection point rightfully remains independent of the initial condition. Thermal strengths exerted by two reservoirs vary as the tugging competition progresses (Table \ref{table2}). All the time the interaction between the hot reservoir and the plate decreases, whereas that between the cold reservoir and the plate increases. However, at $t=700\,s$ (prior to $t_{if}=975\,s$), within a period of $10\,s$, the left side decreases by $1.78\times10^{-3}\,W/K$, and the right side increases by $1.55\times10^{-3}\,W/K$, thus resulting in a net decrease by $2.3\times10^{-4}\,W/K$. At $t=1600\,s$, these roles of influencing are switched, resulting in a net increase by $3\times10^{-5}\,W/K$.
\begin{table}
\begin{center}
\begin{tabular}{ccccc}
\hline
$t\,(s)\,\,\,\,\,\,\,\,\,\,\,$             &hot influence\,\,\,\,\,\,\,\,\,\,\, &cold influence\,\,\,\,\,\,\,\,\,\,\, &$B$ &$B/T$\\
  &$(W/K)$  &$(W/K)$  &$(W/K)$ &$(W/K^{2})$\\\hline
$700\,\,\,\,\,\,\,\,\,\,\,$             &$0.40281\,\,\,\,\,\,\,\,\,\,\,$ &$0.15258\,\,\,\,\,\,\,\,\,\,\,$ &$0.55539$ &\\\hline
$710\,\,\,\,\,\,\,\,\,\,\,$             &$0.40103\,\,\,\,\,\,\,\,\,\,\,$ &$0.15413\,\,\,\,\,\,\,\,\,\,\,$ &$0.55516$ &\\\hline
$1600\,\,\,\,\,\,\,\,\,\,\,$             &$0.34055\,\,\,\,\,\,\,\,\,\,\,$ &$0.21404\,\,\,\,\,\,\,\,\,\,\,$ &$0.55458$ &\\\hline
$1610\,\,\,\,\,\,\,\,\,\,\,$             &$0.34037\,\,\,\,\,\,\,\,\,\,\,$ &$0.21423\,\,\,\,\,\,\,\,\,\,\,$ &$0.55461$ &\\\hline
\end{tabular}
\caption{\textbf{Influences on entropy variation rates, $\bm B$, for the universe consisting of one hot reservoir, one cold reservoir, and an aluminum-plate system.} ``Hot influence'' is the abbreviation for ``thermal influence of hot reservoir on the magnitude of $B(t)$'', defined by the first term in the right-hand side of Eq. $(4)$.}
\label{table2}
\end{center}
\end{table}
This division point constitutes the entropy inflection point, indicated by Eqs. (\ref{6}) and (\ref{7}). Prior to this division, $S_{a}$ is decelerating, whereas afterwards it is accelerating.
The degree of conspicuousness can be readily increased by increasing $R_{1}$ to, for example, $700\,K$ (Fig. \ref{PIC2}$b$). Although $B$ dips to a minimum and increases afterwards, we have discovered that $B/T$ always monotonically decreases (Supplementary S-$5$), namely,
\begin{equation}\label{9}
d(B/T)/dt\leq0.
\end{equation}\label{9}
To be rigorous, we should write $B_{assembly}/T_{system}$. However, we will omit subscripts hereafter to avoid the clumsiness,. In Fig. \ref{PIC2}$c$, both $B$ and $B/T$ are plotted versus $t$. They are normalized at $t=30\,s$. Physically, the increasing rate of $T$ is greater than that of $B$ is after $t=30\,s$. Hence the ratio, $B/T$, continues to decrease. Under the assumption that the heat transfer between the system and a reservoir can be expressed by $h(R-T)$, Eq. (\ref{9}) is always valid (Table \ref{table2}).\par

Next, let us examine the sensitivity of $B(t)$ to perturbations\cite{18} by suddenly bringing a fan to blow on the right face of the plate, thus increasing the value of $h_{2}$ from $10$ to $40$, at $t=120\,s$ for a duration of $10\,s$. Since the plate initially starts from $R_{2}$, the system's temperature does not deviate much from $R_{2}$ yet at this time. Hence, increasing the value of $h_{2}$ increases the influence only minimally on the magnitude of the second term in Eq. (\ref{4}), thus generating a small burst (Fig. \ref{PIC2}$d$). At $t=1200\,s$, the plate has been warmed to $328.58\,K$, which differs appreciably from $R_{2}=200\,K$. Hence an increase of $h_{2}$ causes a spike in $B$. Similarly, if the system's temperature starts from $R_{1}$ initially, and $h_{1}$ (instead of $h_{2}$) is increased suddenly at $t=1200\,s$, we will observe a spike as well. In summary, Eq. (\ref{4}) helps us to predict when the assembly is vulnerable to external perturbations.\par

Two parameters, $\delta_{1}=B_{ss}-B_{min}$ and $\delta_{2}=T_{ss}-T_{if}$, may be of interest to us. Both of them measure the vulnerability of the entropy accelerating regime. Intuitively, they may vary proportionally, though not necessarily linearly. This intuition is found to be generally correct except for regions indicated by starred curves confined in two dashed lines (Fig. \ref{PIC2}$e$). For example, for $R_{1}=700\,K$, as $\delta_{2}$ increases from $53$ to $82$, $\delta_{1}$ decreases from $0.99$ to $0.58$. Furthermore, as $R_{1}$ increases from $400\,K$ to $700\,K$, both $\delta_{1}$ and $\delta_{2}$ increase in general. When $h_{2}$ is weak, e. g. $h_{2}=1$, $\delta_{1}$ increases weakly from $0.004\,(R_{1}=400\,K)$ to $0.055\,(R_{1}=700\,K)$. When $h_{2}$ is moderately strong, e. g. $h_{2}=22$, $\delta_{1}$ increases significantly from $0.0950\,(R_{1}=400\,K)$ to $0.9945\,(R_{1}=700\,K)$, suggesting that bringing a fan to blow the cold-reservoir air is more ``destructive'' (causing larger values of $B$) than to blow the hot-reservoir air. Finally, from Fig. \ref{PIC2}$d$, it appears that all values of $\delta_{1}$, are positive. As a matter of fact, it can be analytically proved that indeed they always are (Supplementary S-$6$). It is surprising to discover that, when the independent variable in the abscissa is non-dimensionalized into $(T_{ss}-T_{if})/T_{if}$ and the dependent variable in $y$ axis to $(B_{ss}-B_{min})/B_{min}$, suddenly all curves merge into a straight line, with a slope being equal to $0.5$. In other words, we obtain an entropy inflection relation
\begin{equation}\label{10}
T_{ss}/T_{if}=2B_{ss}/B_{min}-1
\end{equation}
for $n\geq2$, regardless of values of reservoir temperatures and heat transfer coefficients. (Supplementary S-$7$). Incidentally, for $n=1$, we know $B_{ss}=B_{min}=0$ and $T_{ss}/T_{if}=1$. Since $r_{T}=2r_{B}-1$, where $r_{T}=T_{ss}/T_{if}$ and $r_{B}=B_{ss}/B_{min}$, we can readily obtain $r_{T}-r_{B}=r_{B}-1$, which is greater than $0$ according to Supplementary S-$6$. Therefore, $r_{T}>r_{B}$, suggesting that $r_{B}$ is always dwarfed by $r_{T}$, and that it cannot possibly grow explosively, unless $T_{ss}$ is nonexistent or very high. The relation tempts us to conjecture that, for example, upon the onset of hurricanes, the atmosphere is experiencing drastic changes, and no longer behaves as a regular reservoir, thus preventing the steady state of the system from existing. In Fig. \ref{PIC2}$f$, the vulnerability, $\delta_{1}$, is observed to peak in the vicinity of $h_{2}/h_{1}\approx2$, and diminish rapidly as the ratio decreases, exhibiting an asymmetrical trend. At $h_{2}/h_{1}=10$ and $R_{1}=700\,K$, we observe that $\delta_{1}$ is approximately $0.6\,W/K$, implying that, even if $h_{1}$ and $h_{2}$ vary appreciably in Newton's cooling law (Eq. (\ref{1})), $B_{ss}-B_{min}>0$ remains to be a fact.\par

Instead of an aluminum plate, let us now bring an aluminum triangular block, and immerse it in three reservoirs simultaneously (Fig. \ref{PIC2}$g$). Again, the inflection point, being inconspicuous, can be only detected in the enlarged inset on the purple line (Fig. \ref{PIC2}$h$). The final steady-state value, $B_{ss}$, is derived as
\begin{equation}\label{11}
B_{ss}=\displaystyle{\sum_{i=1}^{3}}h_{i}A_{i}\left(\frac{T_{ss}}{R_{i}}+\frac{R_{i}}{T_{ss}}-2\right),
\end{equation}
where $T_{ss}=(\displaystyle{\sum_{i=1}^{3}}h_{i}A_{i}R_{i})\Big/(\displaystyle{\sum_{i=1}^{3}}h_{i}A_{i})$, and is computed to be $0.6420\,W/K$. Other three curves represent cases in which only a single reservoir interacts with the system. Inflection points are absent in all of them. Values of $B$ all diminish to zero, because no heat transfer takes place between the system and the reservoir at the steady state for $n=1$. This contrast suggests that the presence of the inflection point generally leads to large values of $B(t)$, forcing the universe to lie in the entropy accelerating regime for long periods of times.\par
\begin{figure}[h]
\subfigure{\includegraphics[width=0.23\textwidth]{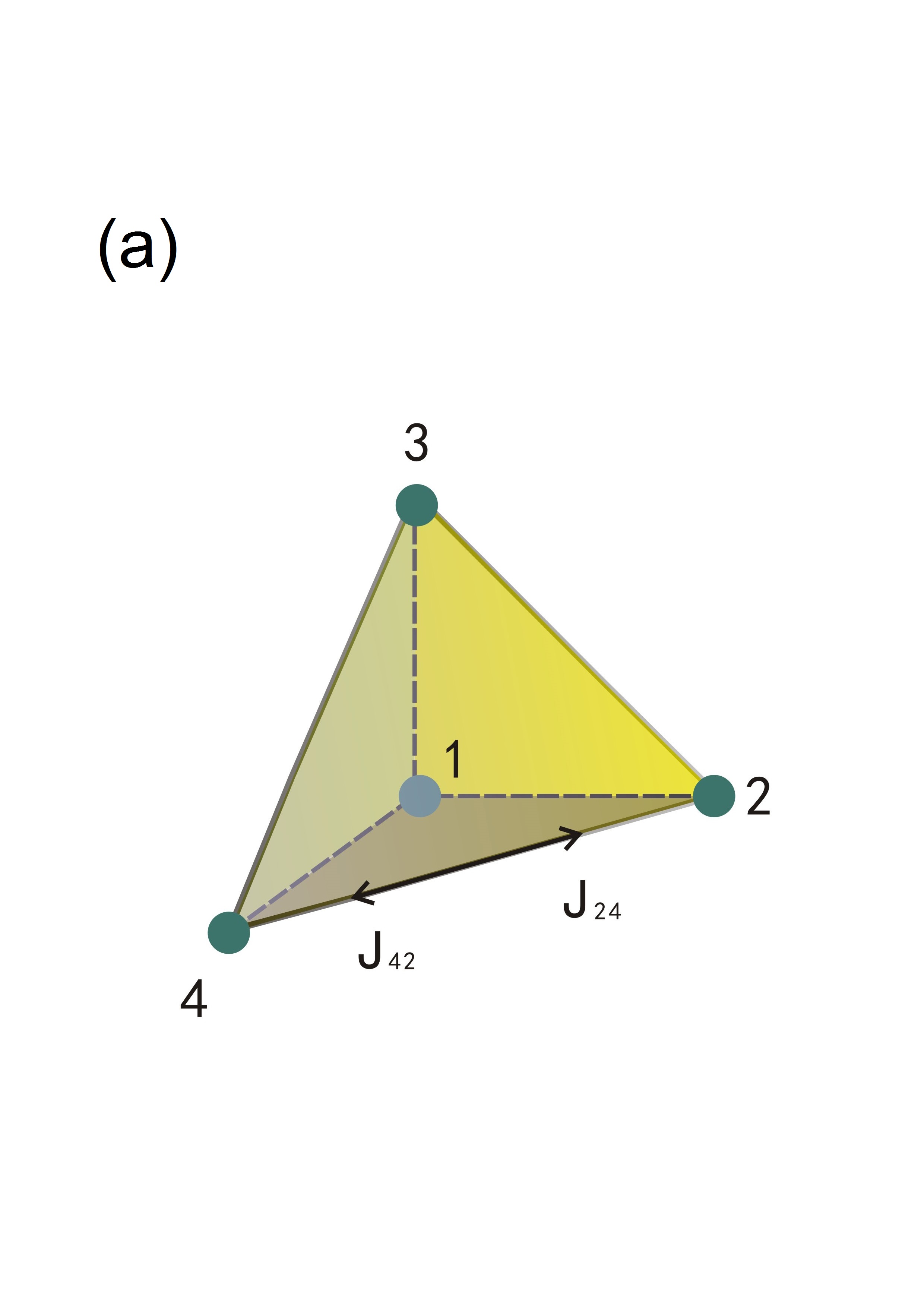}}
\subfigure{\includegraphics[width=0.23\textwidth]{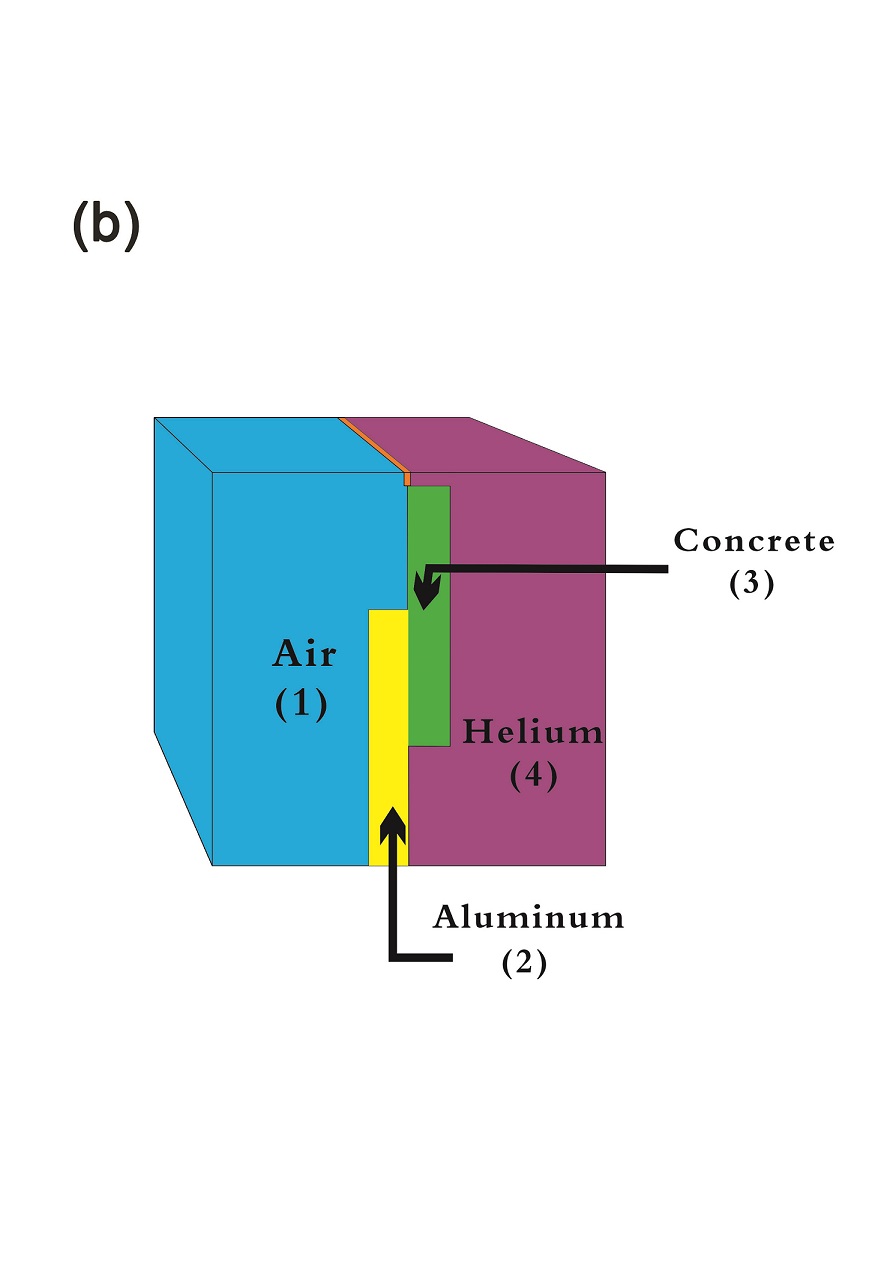}}
\caption{Four-node universe and a physical example. $\bm a,$ Nodal representation of the model. Heat flow rates follow the sign convention $J_{ij}=- J_{ji}$. $\bm b,$ A four-node universe consisting of airflow reservoir (node $1$), helium-flow reservoir (node $4$), an aluminum-plate system (node $2$), and a concrete-plate system (node $3$). There also exists a massless membrane dividing two reservoirs, allowing the option of controlling the interaction between them.}
\label{PIC3}
\end{figure}
Finally, we present an idealized four-node universe (IFNU), from which seven cases are generated. (Figs. \ref{PIC3}$a$, $b$) ``Four'' is the minimum number to remain sufficiently generalized by comprising both multiple systems ($2$) and multiple reservoirs ($2$). Clearly, previous three examples qualify as special cases of the IFNU model with $1R/1s$, $2R/1s$, and $3R/1s$, respectively. Four major idealizations have been made: $(1)$ All processes are assumed to be internally reversible. The adverb ``internally'' is referred to the system, and will be slightly elaborated. A can of beer ($m=0.25\,kg,\,c_{v}=4180\,J/kgK$) at $5^{\circ}C$ is taken out from the refrigerator to be warmed up in a room at $25^{\circ}C$. It is then taken back into the refrigerator. These warming and cooling processes are completely reversible to each other. After this cycle, nothing has changed to the beer. However, the universe has changed irreversibly, because the thermal energy of $20900\,J$ ($0.25\times4180\times20$) has forever transferred from the room to the refrigerator. Irreversibility has been comprehensively reviewed\cite{19}, and rigorously studied within the framework of quantum mechanics\cite{20}, theory of entanglement\cite{21}, and experimental verification of Landauer's erasure principle\cite{22}.
$(2)$ The interaction between node $i$ and node $j$ is exclusively expressed by Newton's law of cooling, namely, $J_{ij}=A_{ij}h_{ij}(T_{j}-T_{i})$, where $h_{ij}$ is assumed to be a constant. The justification of this assumption will be further explained in Methods. $(3)$ The model is isolated such that no heat transfer or mass transfer crosses the model's boundary. $(4)$ Inside a node itself, both thermal equilibrium\cite{23} and mechanical equilibrium prevail. Namely, the temperature and the pressure must be uniform. If a $1D$ fin, whose base temperature is $100^{\circ}C$, is immersed in a reservoir at $25^{\circ}C$, the temperature distribution inside the fin is clearly a function of $x$. Under this circumstance, the fin can be approximated as an assembly of multiple segments (systems) whose temperatures are uniform individually. As the Carnot cycle is an idealized heat engine and serves as a limiting case for real ones, the IFNU model is meant to serve as a limiting case. Temperatures and entropy variation rates of four bodies are theoretically derived (supplementary S-$8$). Numerical simulations using COMSOL\cite{24, 25} have been conducted to retrieve appropriate values of heat transfer coefficients, which are in close agreement with classical literature values\cite{26}.\par
\begin{table}
\footnotesize
\begin{center}
\begin{tabular}{ccccc}
\hline
Case              &Description of    &$T_{if}$  &$B_{min}$  &$T_{ss}$ \\
             &the universe   &($K$)   &($W/K$)   &($K$)  \\
$1$ (standard)    &$2R/2s$   &$324.26$ &$0.1848$ &$339.97$ \\ \hline
$2$               &$2R/2s$   &$324.26$  &$0.2681$  &$n/a$ \\
               &$h_{14}=100\neq0$   &  & \\ \hline
$3$               &$2R/2s$   &$336.62$ &$0.1709$ &$339.97$ \\
               &$h_{23}=0$   &  & \\\hline
$4$               &$2R/2s$   &$325.48$ &$0.2681$ &$339.97$ \\
               &$h_{12}=10(1+e^{-0.02t})$   & & \\\hline
$5$               &$1R/3s$   &$n/a$ &$0$ &$400$ \\
               &$m_{4}=100$   & &  \\\hline
$6$               &$0R/4s$   &$n/a$ &$0$ &$318.68$ \\
               &$m_{1}=100,\,m_{4}=100$   & & \\\hline
$7$               &$3R/1s$   &qualitative &description &only \\\hline
\end{tabular}
\caption{\textbf{Seven case studies of an IFNU given in Fig. \ref{PIC3}.} Case $1$ is the standard one. All other cases deviate from Case $1$ by only one condition indicated.}
\label{table3}
\end{center}
\end{table}

\hspace{-3mm}\textbf{Methods}\par

The discovery of entropy inflection points is deliberately made by analytical methods such that results are immune from numerical finite-difference truncation errors. The Crank-Nicholson time integration algorithm\cite{27,28} of second-order $(\Delta t)^{2}$ accuracy is used primarily for the purpose of validating the correctness of algebraic manipulations in the analytical method. When $\Delta t=1\,s$ is taken, the theoretical result and the numerical result are observed to be indistinguishable, rendering it unnecessary to adopt schemes of higher-order accuracies. The commercial code, COMSOL, whose discretization algorithm is based on the Galerkin finite element method\cite{29,30}, is used to determine appropriate values of the heat transfer coefficient, $h$. Governing equations for the mass conservation, three-direction Navier-Stokes, the energy transport, and the equation of state are written as
\begin{equation}
\frac{\partial \rho}{\partial t}+\nabla\cdot\left(\rho\bm{\upsilon}\right)=0,\nonumber
\end{equation}
\begin{equation}
\frac{\partial}{\partial t}\left(\rho\bm{\upsilon}\right)+\rho\bm{\upsilon}\cdot\left(\nabla\bm{\upsilon}\right)=\mu\nabla^{2}\bm{\upsilon}-\nabla p+\rho \bm{g},\nonumber
\end{equation}
\begin{equation}
\frac{\partial}{\partial t}\left(\rho c_{v}T\right)+\rho c_{p}\bm{\upsilon}\cdot\left(\nabla T\right)=\kappa\nabla^{2}T,\nonumber
\end{equation}
and
\begin{equation}
p=\rho RT,\nonumber
\end{equation}
where $\bm{g}=-g\vec{j}$ and $g$ is the gravitational acceleration. In general, an appropriate heat transfer coefficient for steady-state flow problems can be readily determined by
\begin{equation}
h=-\kappa(\frac{\partial T}{\partial x})_{x=0}\Big/(T_{s}-T_{\infty}),\nonumber
\end{equation}
where $T_{s}$ is the surface temperature of a vertical plate, which is exposed to a fluid flow at $T_{\infty}$  on its right. Since problems considered in the present study are mostly time-dependent, the validity of the Newton's law of cooling, Eq. (\ref{1}), needs to be examined with care. Within one second, the aluminum plate having a thickness of $0.005\,m$ increases its temperature by approximately $0.23^{\circ}C$ when its surface is exposed to two reservoir at $500\,K$ and $200\,K$ (Supplementary S-$2$). The average streamwise velocity in typical free-convection airflows lies in the order of magnitude of $1\,m/s$. For a plate of $1\,m$ in length, fluid particles thus experience a mere change of $0.23^{\circ}C$ in $T_{s}$ while traveling through the system domain. Consequently, this flow phenomenon can be reasonably assumed to be in quasi-steady state, which means that the heat conduction in the plate is transient while the heat convection in the flow is steady. In addition, from Fig. \ref{PIC2}$d$, it is observed that the vulnerability, $\delta_{2}= T_{ss}-T_{if}$, can become as large as $34^{\circ}C$ when $R_{1}=500\,K$ and $R_{2}=200\,K$. Hence it is safe to assume that the existence of entropy inflection points is not caused by the assumption of the Newton's law of cooling.

\hspace{-3mm}\textbf{Supplementary Information} is linked to the online version of the paper at www.nature.com/nature.\par

\hspace{-3mm}\textbf{Acknowledgments} This work is supported in part by the $863$ project of China under Grant $2013AA03A107$, Major Science and Technology Project between University-Industry Cooperation in Fujian Province under Grant $Nos.\,2011H6025$ and $2013H6024$, NNSF of China under Grant $11104230$, Key Project of Fujian Province under Grant $2012H0039$ and the Institute for Complex Adaptive Matter, University of California, Davis, under Grant $ICAM-UCD1308291$.\par

\hspace{-3.6mm}\textbf{Author Contributions} T. M. S. and Z. J. G. conceived the project. H. M., L. R., P. J. P., and Z. C. provided technical supports.\par

\hspace{-3.6mm}\textbf{Author Information} Reprints and permission information is available at www.nature.com/reprints. Authors declare no competing financial interests. Readers are welcome to comment on the online version of this article at www.nature.com/nature. Correspondence and requests for materials should be addressed to T. M. S. (tmshih@xmu.edu.cn) or Z. C. (chenz@xmu.edu.cn).\par

\clearpage

\hspace{-3.6mm}\textbf{Supplementary information}\par

\hspace{-3.6mm}\textbf{S-$1$ Definition of disorderliness for microscopic systems of particles}\label{Definition of disorderliness for microscopic systems of particles}\par

Let us assign $1$ to black marbles and $0$ to white marbles, and define the grayness $S_{u}(t)$ of the system at time $t$ as
\begin{equation}
S_{u}(t)=\displaystyle{\sum_{<i,j>}}\left(x_{i}-x_{j}\right)^{2},\,\,\,\,\,\,\,\,\,\,\,\,\,\,\,\,\,\,\,\,\,\,\,\,\,\,\,\,\,\,\,\,\,\,\,\,\,\,\,\,\,\,\,\,\,\,\,\,\,\,\,\,\,\,\,\,\,\,\,\,(S-1)\nonumber
\end{equation}
where $<>$ presents nearest neighbors, and $x_{i}$, with $i\in\left[1,N\right]$, is the number associated with the $i^{th}$ site. The initial state is prescribed for $i\in\left[1,N/2\right],\,x_{i}=1$, and $i\in\left[N/2+1,N\right],\,x_{i}=0$. Each site $i$ is given with an interval $\left[(i-1)/N,i/N\right)$. The time evolution of our system is governed by the following rule: at each time step $t$, we generate two random numbers $y_{1}$ and $y_{2}$, where $y_{1},y_{2}\in\left[0,1\right]$. Then if $y_{1}\in\left[(i-1)/N,i/N\right]$ and $y_{2}\in\left[(j-1)/N,j/N\right]$, we exchange values of site $i$ and site $j$. The variation rate of grayness is defined as
\begin{equation}
B=\left[S_{u}(t)-S_{u}\left(t-\Delta t\right)\right]/\Delta t.\,\,\,\,\,\,\,\,\,\,\,\,\,\,\,\,\,\,\,\,\,\,\,\,\,\,\,\,\,\,\,\,\,\,\,\,\,\,\,\,\,\,\,\,\,\,\,(S-2)\nonumber
\end{equation}

\hspace{-3.6mm}\textbf{S-$2$ Validity of the lumped-capacitance model for a vertical aluminum plate immersed in two reservoirs}\label{Validity of the lumped-capacitance model for a vertical aluminum plate immersed in two reservoirs}\par
In reference to Fig. $2\,a$, the first law of thermodynamics applied to a segment of the control volume containing node $i$ can be written as
\begin{equation}
\left(\Delta m\right)c_{v}\left(T_{i}-T_{i}^{p}\right)=\left(J_{i,w}-J_{i,e}\right)\Delta t,\,\,\,\,\,\,\,\,\,\,\,\,\,\,\,\,\,\,\,\,\,\,\,\,\,\,\,\,\,(S-3)\nonumber
\end{equation}
where $\Delta m=\rho A\Delta x$; the subscript ``p'' denotes ``at the previous time step''; the subscript ``w'' denotes ``west to $i$''; the subscript ``e'' denotes ``east to $i$''. Also, according to Fourier's law and the Crank-Nicholson algorithm, we obtain
\begin{equation}
J_{i,w}=0.5\kappa A\left(\frac{T_{i-1}-T_{i}}{\Delta x}+\frac{T_{i-1}^{p}-T_{i}^{p}}{\Delta x}\right),\nonumber
\end{equation}
and
\begin{equation}
J_{i,e}=0.5\kappa A\left(\frac{T_{i}-T_{i+1}}{\Delta x}+\frac{T_{i}^{p}-T_{i+1}^{p}}{\Delta x}\right),\nonumber
\end{equation}
which can be substituted into Eq. ($S-3$) to yield
\begin{equation}
\begin{split}
&a(i,i-1)T_{i-1}+a(i,i)T_{i}+a(i,i+1)T_{i+1}=b(i),\\
&\text{for }i=2,3,\cdots,nx,\,\,\,\,\,\,\,\,\,\,\,\,\,\,\,\,\,\,\,\,\,\,\,\,\,\,\,\,\,\,\,\,\,\,\,\,\,\,\,\,\,\,\,\,\,\,\,\,\,\,\,\,\,\,\,\,\,\,\,\,\,\,\,\,\,\,\,\,\,\,\,\,\,\,(S-4)\nonumber
\end{split}
\end{equation}
where $a(i,i-1)=-0.5r$, $a(i,i)=r+1$, $a(i,i+1)=-0.5r$, $b(i)=T_{i}^{p}+e_{i}$, and $e_{i}=0.5r(T_{i-1}^{p}+T_{i+1}^{p})-rT_{i}^{p}$.
Energy balance over the boundary segments containing $1$ and $nx+1$ can be conducted similarly, except that the mass is $0.5\Delta m$ for either node. Corresponding coefficients can be derived to be
\begin{equation}
\begin{split}
&a(1,1)=0.5+0.5(r+c_{1});\,a(1,2)=-0.5r;\\
&b(1)=0.5T_{1}^{p}+0.5c_{1}R_{1}+e_{1};\,a(nx+1,nx)=-0.5r;\\
&a(nx+1,nx+1)=0.5+0.5(r+c_{1});\,\\
&b(nx+1)=0.5T_{nx+1}^{p}+0.5c_{1}R_{2}+e_{L};\nonumber
\end{split}
\end{equation}
where
\begin{equation}
\begin{split}
&r=\alpha\Delta t/(\Delta x)^{2};\,\alpha=\kappa/(\rho c_{v});\,b_{1}=h\Delta x/\kappa;\,c_{1}=r\times b_{1};\\
&e_{1}=0.5c_{1}\left(R_{1}-T_{1}^{p}\right)-0.5r\left(T_{1}^{p}-T_{2}^{p}\right);\\
&e_{L}=-0.5c_{1}\left(T_{nx+1}^{p}-R_{2}\right)+0.5r\left(T_{nx}^{p}-T_{nx+1}^{p}\right);\nonumber
\end{split}
\end{equation}
Numerical results for $\Delta t=1\,s$ show that $(T(1)-T(nx+1))/T(1)\approx1.57\times10^{-4}$ under $R_{1}=500\,K$ and $R_{2}=200\,K$ within the beginning $5\,s$ averagely. This minute $\Delta T$ value suggests that the assumption of a uniform $T(x)$ within the aluminum plate is valid. Also, during this period of time, we check the global energy balance, which states that both the net energy entering the plate and the internal-energy increase of the plate are equal to $1.49\times10^{4}\,J$.

\hspace{-3.6mm}\textbf{S-$3$ Derivations of Eqs.$(4,5)$}\label{Derivation of Eq.(4)}\par
Let there be two reservoirs at temperatures $R_{1}$ and $R_{2}$, as well as one system at $T$. If there are two or more systems, the algebraic procedure becomes more complicated, but the essential concept remains the same.\par

The infinitesimal entropy variation rate of the universe can be written as
\begin{equation}
dS_{u}=\frac{-\delta Q_{1}}{R_{1}}+\frac{\delta Q_{1}-\delta Q_{2}}{T}+\frac{\delta Q_{2}}{R_{2}},\nonumber
\end{equation}
or $dS_{u}=\delta Q_{1}\left(1/T-1/R_{1}\right)+\delta Q_{2}\left(1/R_{2}-1/T\right)$,\\
where $\delta Q_{1}=A_{1}h_{1}\left(R_{1}-T\right)dt$ and $\delta Q_{2}=A_{2}h_{2}\left(T-R_{2}\right)$. Hence,
\begin{equation}
B=A_{1}h_{1}\left(\frac{R_{1}}{T}+\frac{T}{R_{1}}-2\right)+A_{2}h_{2}\left(\frac{R_{2}}{T}+\frac{T}{R_{2}}-2\right),\nonumber
\end{equation}
which can be differentiated once to yield the entropy acceleration as
\begin{equation}
\frac{d^{2}S_{u}}{dt^{2}}=\left[A_{1}h_{1}\left(-\frac{R_{1}}{T^{2}}+\frac{1}{R_{1}}\right)+A_{2}h_{2}\left(-\frac{R_{2}}{T^{2}}+\frac{1}{R_{2}}\right)\right]\frac{dT}{dt}.\nonumber
\end{equation}

\hspace{-3.6mm}\textbf{S-$4$ Proof of $T_{if}\leq T_{ss}$}\label{Proof of $T_{infl}<T_{ss}$}\par

Without the loss of generality, let us consider the existence of $3$ reservoirs and absorb $A_{1}$, $A_{2}$, and $A_{3}$ into $h_{1}$, $h_{2}$, and $h_{3}$, respectively, in the derivation. Since $h_{1}$, $h_{2}$, $h_{3}$, $R_{1}$, $R_{2}$, and $R_{3}$ are all positive numbers, we obtain
\begin{equation}
\begin{split}
&\frac{h_{1}h_{2}}{R_{1}R_{2}}\left(R_{1}^{2}+R_{2}^{2}-2R_{1}R_{2}\right)+\frac{h_{2}h_{3}}{R_{2}R_{3}}\left(R_{2}^{2}+R_{3}^{2}-2R_{2}R_{3}\right)\\
&+\frac{h_{3}h_{1}}{R_{3}R_{1}}\left(R_{3}^{2}+R_{1}^{2}-2R_{3}R_{1}\right)\geq0,\nonumber
\end{split}
\end{equation}
which, when straightforwardly expanded into nine terms, becomes
\begin{equation}
\begin{split}
&\frac{h_{1}h_{2}R_{1}}{R_{2}}+\frac{h_{1}h_{2}R_{2}}{R_{1}}-2h_{1}h_{2}+\frac{h_{2}h_{3}R_{2}}{R_{3}}+\frac{h_{2}h_{3}R_{3}}{R_{2}}-2h_{2}h_{3}\\
&+\frac{h_{3}h_{1}R_{3}}{R_{1}}+\frac{h_{3}h_{1}R_{1}}{R_{3}}-2h_{3}h_{1}\geq0.\nonumber
\end{split}
\end{equation}
Another trivial rearrangement by inserting $h_{1}^{2}+h_{2}^{2}+h_{3}^{2}-h_{1}^{2}-h_{2}^{2}-h_{3}^{2}$ leads to
\begin{equation}
\begin{split}
&h_{1}^{2}+h_{2}^{2}+h_{3}^{2}+\frac{h_{1}h_{2}R_{1}}{R_{2}}+\frac{h_{1}h_{2}R_{2}}{R_{1}}+\frac{h_{2}h_{3}R_{2}}{R_{3}}+\frac{h_{2}h_{3}R_{3}}{R_{2}}\\
&+\frac{h_{3}h_{1}R_{1}}{R_{3}}+\frac{h_{3}h_{1}R_{3}}{R_{1}}-2h_{1}h_{2}-2h_{2}h_{3}-2h_{1}h_{3}-h_{1}^{2}\\
&-h_{2}^{2}-h_{3}^{2}\geq0,\nonumber
\end{split}
\end{equation}
which can be readily cast into
\begin{equation}
\begin{split}
&\left(h_{1}R_{1}+h_{2}R_{2}+h_{3}R_{3}\right)\left(\frac{h_{1}}{R_{1}}+\frac{h_{2}}{R_{2}}+\frac{h_{3}}{R_{3}}\right)\\
&-\left(h_{1}+h_{2}+h_{3}\right)^{2}\geq0,\nonumber
\end{split}
\end{equation}
or
\begin{equation}
\frac{h_{1}R_{1}+h_{2}R_{2}+h_{3}R_{3}}{\left(h_{1}+h_{2}+h_{3}\right)^{2}}-\frac{1}{\left(h_{1}/R_{1}+h_{2}/R_{2}+h_{3}/R_{3}\right)}\geq0,\nonumber
\end{equation}
or
\begin{equation}
\frac{\left(h_{1}R_{1}+h_{2}R_{2}+h_{3}R_{3}\right)^{2}}{\left(h_{1}+h_{2}+h_{3}\right)^{2}}-\frac{h_{1}R_{1}+h_{2}R_{2}+h_{3}R_{3}}{h_{1}/R_{1}+h_{2}/R_{2}+h_{3}/R_{3}}\geq0,\nonumber
\end{equation}
or with a reasonable extension to
\begin{equation}
\frac{\displaystyle{\sum_{i=1}^{n}}\left(h_{i}R_{i}\right)^{2}}{(\displaystyle{\sum_{i=1}^{n}}h_{i})^{2}}-\frac{\displaystyle{\sum_{i=1}^{n}}\left(h_{i}R_{i}\right)}{\displaystyle{\sum_{i=1}^{n}}\left(h_{i}/R_{i}\right)}\geq0.\nonumber
\end{equation}
Finally, we obtain
\begin{equation}
T_{ss}^{2}-T_{if}^{2}\geq0,\text{ or }T_{if}\leq T_{ss}.\nonumber
\end{equation}\par

\hspace{-3.6mm}\textbf{S-$5$ Proof of $B/T$ always decreases}\label{Proof of}\par
Since $B=h_{1}(R_{1}/T+T/R_{1}-2)+h_{2}(R_{2}/T+T/R_{2}-2)$,
we obtain
\begin{equation}
Y=B/T=\left(\frac{h_{1}}{R_{1}}+\frac{h_{2}}{R_{2}}\right)
+\left(\frac{R_{1}h_{1}}{T^{2}}+\frac{R_{2}h_{2}}{T^{2}}\right)-\frac{2}{T}(h_{1}+h_{2}),\nonumber
\end{equation}
Or
\begin{equation}
\frac{dY}{dt}=\lambda_{1}\frac{d}{dt}(T^{-2})
-2\lambda_{3}\frac{d}{dt}(T^{-1}),\nonumber
\end{equation}
where $\lambda_{1}=h_{1}R_{1}+h_{2}R_{2}$ and $\lambda_{3}=h_{1}+h_{2}$. Further manipulations yield
\begin{equation}
\frac{dY}{dt}=\frac{2}{T^{2}}\frac{dT}{dt}\left(\lambda_{3}-\frac{\lambda_{1}}{T}\right),\nonumber
\end{equation}
which can be rearranged to become
\begin{equation}
\frac{dY}{dt}=-(\frac{2}{T^{3}})(\frac{dT}{dt})[h_{1}(R_{1}-T)+h_{2}(R_{2}-T)].\nonumber
\end{equation}
For heating processes, $dT/dt$ is positive, and the quantity inside the brackets, representing the net energy entering the system, should also be positive. Hence, $dY/dt$ should be positive. For cooling processes, the argument is similar.

\hspace{-3.6mm}\textbf{S-$6$ Proof of $B_{min}\leq B_{ss}$}\label{Proof of}\par
For conivenience, let us introduce $\lambda_{1}=h_{1}R_{1}+h_{2}R_{2}$ and $\beta_{1}=h_{1}R_{2}+h_{2}R_{1}$. Then we start with
\begin{equation}
R_{1}^{2}+R_{2}^{2}\geq2R_{1}R_{2},\nonumber
\end{equation}
which can be trivially manipulated into
\begin{equation}
\lambda_{1}\beta_{1}\geq\left(h_{1}+h_{2}\right)^{2}\left(R_{1}R_{2}\right),\nonumber
\end{equation}
or
\begin{equation}
\lambda_{1}^{1/2}\beta_{1}^{1/2}-\left(h_{1}+h_{2}\right)\left(R_{1}R_{2}\right)^{1/2}\geq0,\,\,\,\,\,\,\,\,\,\,\,\,\,\,\,\,\,\,\,\,\,\,\,\,\,\,\,\,\,\,(S-5)\nonumber
\end{equation}
as long as $h_{1}$ and $h_{2}$ are positive numbers. Without loss of generality, let us neglect writing $A_{1}$ and $A_{2}$ for the reason that they always appear adjacently to $h_{1}$ and $h_{2}$. In the event that $A_{1}\neq A_{2}$, we can simply replace $h_{1}$ and $h_{2}$ with $A_{1}h_{1}$ and $A_{2}h_{2}$. Hence, from Eq. ($4$), we can obtain
\begin{equation}
B_{ss}-B_{min}=\left(T_{ss}-T_{if}\right)\left(\frac{\beta_{1}}{R_{1}R_{2}}-\frac{\lambda_{1}}{T_{if}T_{ss}}\right).\,\,\,\,\,\,(S-6)\nonumber
\end{equation}
According to Eq. (\ref{7}) and the fact that
\begin{equation}
T_{ss}=\lambda_{1}/\left(h_{1}+h_{2}\right),\,\,\,\,\,\,\,\,\,\,\,\,\,\,\,\,\,\,\,\,\,\,\,\,\,\,\,\,\,\,\,\,\,\,\,\,\,\,\,\,\,\,\,\,\,\,\,\,\,\,\,\,\,\,\,\,\,\,\,\,\,\,(S-7)\nonumber
\end{equation}
the product of $T_{ss}$ and $T_{if}$ becomes
\begin{equation}
T_{ss}T_{if}=\frac{\lambda_{1}^{3/2}\left(R_{1}R_{2}\right)^{1/2}}{\left(h_{1}+h_{2}\right)\beta_{1}^{1/2}},\,\,\,\,\,\,\,\,\,\,\,\,\,\,\,\,\,\,\,\,\,\,\,\,\,\,\,\,\,\,\,\,\,\,\,\,\,\,\,\,\,\,\,\,\,\,\,\,\,\,\,\,(S-8)\nonumber
\end{equation}
which can be substituted into Eq. ($S-6$) to offer us
\begin{equation}
\begin{split}
&B_{ss}-B_{min}\,\,\,\,\,\,\,\,\,\,\,\,\,\,\,\,\,\,\,\,\,\,\,\,\,\,\,\,\,\,\,\,\,\,\,\,\,\,\,\,\,\,\,\,\,\,\,\,\,\,\,\,\,\,\,\,\,\,\,\,\,\,\,\,\,\,\,\,\,\,\,\,\,\,\,\,\,\,\,\,\,\,\,\,\,\,(S-9)\\
=&\frac{\left(T_{ss}-T_{if}\right)\beta_{1}^{1/2}}{\left(R_{1}R_{2}\right)\lambda_{1}^{1/2}}\left[(\lambda_{1}\beta_{1})^{1/2}-\left(h_{1}+h_{2}\right)(R_{1}R_{2})^{1/2}\right].\nonumber
\end{split}
\end{equation}
Based on Eqs. (\ref{8}) and ($S-5$), we are now able to assert
\begin{equation}
B_{min}\leq B_{ss}\,\,\,\,\,\text{always.}\nonumber
\end{equation}

\hspace{-3.6mm}\textbf{S-$7$ Proof of $\left(T_{ss}-T_{if}\right)/T_{if}=2\left(B_{ss}-B_{min}\right)/B_{min}$}\label{Proof}\par
For convenience of algebraic manipulations, let us define $\lambda_{1}=\displaystyle{\sum_{i=1}^{n}}h_{i}R_{i}$, $\lambda_{2}=\displaystyle{\sum_{i=1}^{n}}h_{i}/R_{i}$, and $\lambda_{3}=\displaystyle{\sum_{i=1}^{n}}h_{i}$. Thus, we obtain $T_{ss}=\lambda_{1}/\lambda_{3}$, $T_{if}=(\lambda_{1}/\lambda_{2})^{1/2}$, and naturally
\begin{equation}
T_{if}T_{ss}=\frac{\lambda_{1}^{3/2}}{\lambda_{2}^{1/2}\lambda_{3}}.\,\,\,\,\,\,\,\,\,\,\,\,\,\,\,\,\,\,\,\,\,\,\,\,\,\,\,\,\,\,\,\,\,\,\,\,\,\,\,\,\,\,\,\,\,\,\,\,\,\,\,\,\,\,\,\,\,\,\,\,\,\,\,\,\,\,\,\,\,\,\,\,\,\,\,\,\,\,\,(S-10)\nonumber
\end{equation}
Based on Eq. (\ref{4}), we obtain (let $A_{i}$ be absorbed in $h_{i}$ for clarity)
\begin{equation}
B=\displaystyle{\sum_{i=1}^{n}}h_{i}\left(\frac{T}{R_{i}}+\frac{R_{i}}{T}-2\right),\,\,\,\,\,\,\,\,\,\,\,\,\,\,\,\,\,\,\,\,\,\,\,\,\,\,\,\,\,\,\,\,\,\,\,\,\,\,\,\,\,\,\,\,\,\,\,\,\,\,\,\,(S-11)\nonumber
\end{equation}
which yields
\begin{equation}
B_{ss}=T_{ss}\lambda_{2}+\frac{\lambda_{1}}{T_{ss}}-2\lambda_{3}\,\,\,\,\,\,\,\,\,\,\,\,\,\,\,\,\,\,\,\,\,\,\,\,\,\,\,\,\,\,\,\,\,\,\,\,\,\,\,\,\,\,\,\,\,\,\,\,\,\,\,\,\,\,\,\,\,\,\,\,\,(S-12)\nonumber
\end{equation}
and
\begin{equation}
B_{min}=T_{if}\lambda_{2}+\frac{\lambda_{1}}{T_{if}}-2\lambda_{3},\,\,\,\,\,\,\,\,\,\,\,\,\,\,\,\,\,\,\,\,\,\,\,\,\,\,\,\,\,\,\,\,\,\,\,\,\,\,\,\,\,\,\,\,\,\,\,\,\,\,\,\,\,(S-13)\nonumber
\end{equation}
leading to
\begin{equation}
B_{ss}-B_{min}=\left(T_{ss}-T_{if}\right)\lambda_{2}-\lambda_{1}\left(\frac{T_{ss}-T_{if}}{T_{ss}T_{if}}\right)\nonumber
\end{equation}
or
\begin{equation}
\frac{B_{ss}-B_{min}}{T_{ss}-T_{if}}=\lambda_{2}-\frac{\lambda_{1}}{T_{ss}T_{if}},\nonumber
\end{equation}
which is the same as Eq. (S-$6$) except that now $n$ is an arbitrary integer. Substitution of Eq. (S-$10$) into the equation above allows us to obtain
\begin{equation}
\frac{B_{ss}-B_{min}}{T_{ss}-T_{if}}=\lambda_{2}-\frac{\lambda_{2}^{1/2}\lambda_{3}}{\lambda_{1}^{1/2}}.\,\,\,\,\,\,\,\,\,\,\,\,\,\,\,\,\,\,\,\,\,\,\,\,\,\,\,\,\,\,\,\,\,\,\,\,\,\,\,\,\,\,\,\,\,\,\,\,\,(S-14)\nonumber
\end{equation}
Next, from Eq. (S-$13$), we also readily obtain
\begin{equation}
\frac{B_{min}}{T_{if}}=\lambda_{2}+\frac{\lambda_{1}}{T_{if}^{2}}-\frac{2\lambda_{3}}{T_{if}}.\,\,\,\,\,\,\,\,\,\,\,\,\,\,\,\,\,\,\,\,\,\,\,\,\,\,\,\,\,\,\,\,\,\,\,\,\,\,\,\,\,\,\,\,\,\,\,\,\,\,\,\,\,\,\,\,\,(S-15)\nonumber
\end{equation}
Upon substitution of the expression for $T_{if}$ into Eq. (S-$15$) produces
\begin{equation}
\frac{B_{min}}{T_{if}}=2\lambda_{2}-\frac{2\lambda_{2}^{1/2}\lambda_{3}}{\lambda_{1}^{1/2}}=2\left(\frac{B_{ss}-B_{min}}{T_{ss}-T_{if}}\right),\,\,\,\,\,\,\,(S-16)\nonumber
\end{equation}
which concludes our proof.

\hspace{-3.6mm}\textbf{S-$8$ Analytical solution of a four-node universe}\label{IFNU model analytical solution}\par
Based on the first law of thermodynamics, the internal energy increase rate of node $1$ is equal to the net heat transfer entering the node (Watts) and the work done to the node. The governing equation for $T_{1}$ can thus be written as
\begin{equation}\label{11}
\begin{split}
m_{1}c_{v_{1}}\frac{dT_{1}}{dt}=&A_{12}h_{12}(T_{2}-T_{1})+A_{13}h_{13}(T_{3}-T_{1})\\
&+A_{14}h_{14}(T_{4}-T_{1})-p_{1}dV_{1},
\end{split}
\end{equation}
where $m_{1}$ denotes the mass of node $1$ ($kg$); $c_{v_{1}}$ the heat capacity ($J/K\cdot kg$); $A_{12}$ the area commonly shared by nodes $1$ and $2$; $h_{12}$ is the heat transfer coefficient between nodes $1$ and $2$; $p_{1}$ the pressure; and $V_{1}$ the volume. Governing equations for nodes $2$, $3$, and $4$ can be written similarly. These four first-order differential equations can be simultaneously solved analytically, subject to specific initial conditions. For example, the final solution can be derived to become
\begin{equation}\label{12}
T_{1}(t)=\xi_{1}\exp(n_{1}t)+\xi_{2}\exp(n_{2}t)+\xi_{3}\exp(n_{3}t)+\xi_{13},
\end{equation}
where values of constants are listed therein. Note that all values of $n_{1}$, $n_{2}$, and $n_{3}$ are negative as expected.\par
By definition, the entropy variation rate for node $1$ can be written as $dS_{1}=\displaystyle{\sum_{j=2}^{4}}\delta Q_{1j}\Big/T_{1}$, or
\begin{equation}\label{13}
\begin{split}
\frac{dS_{1}}{dt}=&\frac{A_{12}h_{12}}{T_{1}}(T_{2}-T_{1})+\frac{A_{13}h_{13}}{T_{1}}(T_{3}-T_{1})\\
&+\frac{A_{14}h_{14}}{T_{1}}(T_{4}-T_{1}).
\end{split}
\end{equation}
A universe consisting of two systems and two reservoirs (Fig. $\ref{PIC3}b$) is considered. Characteristics of this example are: $(a)$ In addition to interactions between systems and reservoirs, both systems also interact with each other. So do both reservoirs. $(b)$ COMSOL\cite{24,25} numerical simulations have been conducted to retrieve appropriate values of heat transfer coefficients, which are then used in entropy-related calculations, rendering the analysis closer to realistic situations. $(c)$ For the entropy analysis, both the analytical method and the numerical method are used to yield indistinguishable results. Relevant conditions and results are listed in Supplementary S-$8$. Case $1$ is the standard scenario, in which node $1$ and node $4$ are reservoirs, while node $2$ and node $3$ are systems. The effective $T_{if}$ of the combined assembly consisting of both node $2$ and node $3$ can be calculated by $T_{eff}=(m_{2} c_{v_{2}}T_{2}+m_{3}c_{v_{3}}T_{3})/(m_{2}c_{v_{2}}+m_{3}c_{v_{3}})$ when $B$ reaches the minimum. All cases $2$ to $6$ deviate from case $1$ slightly. In case $2$, we allow two reservoirs to thermally interact via a nearly massless membrane. Nearly no difference is observed in terms of the existence of an entropy inflection point. However, the steady state cannot be expected because constantly the heat transfer takes place from reservoir $1$ to reservoir $4$. Case $3$ intentionally prohibits two systems from interacting with each other. Due to this prohibiting, $S_{u}$ increases more slowly than that in case $2$ does (according to $B_{min}$ values), suggesting that, hypothetically, the onset of hurricanes might be prevented had we been capable of erecting well-insulated partition walls amid the original epicenter. In case $4$, $h_{12}$ is allowed to gradually decrease from $20$ to $10$, mimicking behaviors observed from the COMSOL simulation results and classical literature values\cite{26}. No alarming difference between this case and the standard case is observed. In case $5$, reservoir $4$ becomes a system with $m_{4}=100\,kg$. No inflection points exist, as mentioned and explained previously. In case $6$, we reduce both $m_{1}$ and $m_{4}$ to $100\,kg$. The IFNU model becomes a $4$-system universe. Again, the claim that no inflection points exist for zero or one reservoir is validated. Case $7$ considers node $2$ as a piston-cylinder system whose volume changes. The head of the cylinder, the circumferential area of the cylinder, and the piston are in contact with reservoirs $1$, $3$, and $4$, respectively. The pressure inside the system is assumed to remain constant. Consequently, applying the first law of thermodynamics to the piston-cylinder system yields $dU=\delta Q-pdV$, which can be readily rearranged into $mc_{p}dT=\delta Q$. Under the assumption of constant pressure, it is seen that the difference between problems of constant volume and those of varying volume becomes minute, except that $c_{v}=717.5\,J/kgK$ should be replaced with $c_{p}=1004.5\,J/kgK$ for air.\par

Let
\begin{equation}
T_{1}(t)=\xi_{1}e^{n_{1}t}+\xi_{2}e^{n_{2}t}+\xi_{3}e^{n_{3}t}+\xi_{13},\nonumber
\end{equation}
\begin{equation}
T_{2}(t)=\xi_{4}e^{n_{1}t}+\xi_{5}e^{n_{2}t}+\xi_{6}e^{n_{3}t}+\xi_{13},\nonumber
\end{equation}
\begin{equation}
T_{3}(t)=\xi_{7}e^{n_{1}t}+\xi_{8}e^{n_{2}t}+\xi_{9}e^{n_{3}t}+\xi_{13},\nonumber
\end{equation}
and
\begin{equation}
T_{4}(t)=\xi_{10}e^{n_{1}t}+\xi_{11}e^{n_{2}t}+\xi_{12}e^{n_{3}t}+\xi_{13},\nonumber
\end{equation}
where $\xi_{i}\,(i=1\sim13)$, $n_{1}$, $n_{2}$, and $n_{3}$ are unknowns ($16$ in total). These nodal temperatures are governed by
\begin{equation}
\frac{dT_{i}}{dt}=\displaystyle{\sum_{j=1}^{4}}c_{ij}T_{j},\,i=1\sim4,\nonumber
\end{equation}
with coefficients satisfying the constraints $\displaystyle{\sum_{j=1}^{4}}c_{ij}=0$ for $i=1\sim4$. After laborious but straightforward algebra by forcing the sum of collected coefficients (in front of each of $e^{n_{1}t}$, $e^{n_{2}t}$, and $e^{n_{3}t}$) to zero and by imposing $4$ initial conditions, we obtain exactly $16$ nonlinear equations, written below as
\begin{equation}
(c_{11}-n_{1})\xi_{1}+c_{12}\xi_{4}+c_{13}\xi_{7}+c_{14}\xi_{10}=0,\,\,\,\,\,\,\,\,\,\,\,\,\,\,\,\,\,(S-17)\nonumber
\end{equation}
\begin{equation}
(c_{11}-n_{2})\xi_{2}+c_{12}\xi_{5}+c_{13}\xi_{8}+c_{14}\xi_{11}=0,\,\,\,\,\,\,\,\,\,\,\,\,\,\,\,\,\,(S-18)\nonumber
\end{equation}
\begin{equation}
(c_{11}-n_{3})\xi_{3}+c_{12}\xi_{6}+c_{13}\xi_{9}+c_{14}\xi_{12}=0,\,\,\,\,\,\,\,\,\,\,\,\,\,\,\,\,\,(S-19)\nonumber
\end{equation}
\begin{equation}
\xi_{1}+\xi_{2}+\xi_{3}+\xi_{13}=T_{o}(1),\,\,\,\,\,\,\,\,\,\,\,\,\,\,\,\,\,\,\,\,\,\,\,\,\,\,\,\,\,\,\,\,\,\,\,\,\,\,\,\,\,\,\,\,\,\,\,\,\,\,\,\,\,\,(S-20)\nonumber
\end{equation}
\begin{equation}
c_{21}\xi_{1}+(c_{22}-n_{1})\xi_{4}+c_{23}\xi_{7}+c_{24}\xi_{10}=0,\,\,\,\,\,\,\,\,\,\,\,\,\,\,\,\,(S-21)\nonumber
\end{equation}
\begin{equation}
c_{21}\xi_{2}+(c_{22}-n_{2})\xi_{5}+c_{23}\xi_{8}+c_{24}\xi_{11}=0,\,\,\,(S-22)\nonumber
\end{equation}
\begin{equation}
c_{21}\xi_{3}+(c_{22}-n_{3})\xi_{6}+c_{23}\xi_{9}+c_{24}\xi_{12}=0,\,\,\,(S-23)\nonumber
\end{equation}
\begin{equation}
\xi_{4}+\xi_{5}+\xi_{6}+\xi_{13}=T_{o}(2),\,\,\,\,\,\,\,\,\,\,\,\,\,\,\,\,\,\,\,\,\,\,\,\,\,\,\,\,\,\,\,\,\,\,\,\,\,\,\,\,\,(S-24)\nonumber
\end{equation}
\begin{equation}
c_{31}\xi_{1}+c_{32}\xi_{4}+(c_{33}-n_{1})\xi_{7}+c_{34}\xi_{10}=0,\,\,\,(S-25)\nonumber
\end{equation}
\begin{equation}
c_{31}\xi_{2}+c_{32}\xi_{5}+(c_{33}-n_{2})\xi_{8}+c_{34}\xi_{11}=0,\,\,\,(S-26)\nonumber
\end{equation}
\begin{equation}
c_{31}\xi_{3}+c_{32}\xi_{6}+(c_{33}-n_{3})\xi_{9}+c_{34}\xi_{12}=0,\,\,\,(S-27)\nonumber
\end{equation}
\begin{equation}
\xi_{7}+\xi_{8}+\xi_{9}+\xi_{13}=T_{o}(3),\,\,\,\,\,\,\,\,\,\,\,\,\,\,\,\,\,\,\,\,\,\,\,\,\,\,\,\,\,\,\,\,\,\,\,\,\,\,\,\,\,(S-28)\nonumber
\end{equation}
\begin{equation}
c_{41}\xi_{1}+c_{42}\xi_{4}+c_{43}\xi_{7}+(c_{44}-n_{1})\xi_{10}=0,\,\,\,(S-29)\nonumber
\end{equation}
\begin{equation}
c_{41}\xi_{2}+c_{42}\xi_{5}+c_{43}\xi_{8}+(c_{44}-n_{2})\xi_{11}=0,\,\,\,(S-30)\nonumber
\end{equation}
\begin{equation}
c_{41}\xi_{3}+c_{42}\xi_{6}+c_{43}\xi_{9}+(c_{44}-n_{3})\xi_{12}=0,\,\,\,(S-31)\nonumber
\end{equation}
and
\begin{equation}
\xi_{10}+\xi_{11}+\xi_{12}+\xi_{13}=T_{o}(4).\,\,\,\,\,\,\,\,\,\,\,\,\,\,\,\,\,\,\,\,\,\,\,\,\,\,\,\,\,\,\,\,\,\,\,(S-32)\nonumber
\end{equation}
For example, consider input data given as\par
$c_{12}=0.10;\,c_{13}=0.23;\,c_{14}=0.31;$\par
$c_{21}=0.12;\,c_{23}=0.37;\,c_{24}=0.14;$\par
$c_{31}=0.24;\,c_{32}=0.15;\,c_{34}=0.16;$\par
$c_{41}=0.29;\,c_{42}=0.16;\,c_{43}=0.18;$\par
$c_{11}=-(c_{12}+c_{13}+c_{14});\,c_{22}=-(c_{21}+c_{23}+c_{24});$\par
$c_{33}=-(c_{31}+c_{32}+c_{34});\,c_{44}=-(c_{41}+c_{42}+c_{43});$\par
and\par
$Tp(1)=1;\,Tp(2)=4;\,Tp(3)=3;\,Tp(4)=2.$\par
\hspace{-3mm}We use the Newton-Raphson method to obtain\par
($\xi_{1}\sim\xi_{13}$)\par
$-0.3998\,\,\,\,\,\,\,\,\,\,0.6487\,\,\,-1.6548$\par
$-0.1891\,\,\,-1.5847\,\,\,\,\,\,\,\,\,\,3.3678$\par
$\,\,\,\,0.1656\,\,\,\,\,\,\,\,\,\,\,0.4300\,\,\,-0.0015$\par
$\,\,\,\,0.3519\,\,\,\,-0.0842\,\,\,-0.6735\,\,\,\,\,\,\,\,\,\,2.4059$\par
$\,\,\,\,\,\,\,\,\,\,\,\,\,\,\,\,\,\,\,\,\,$\par
($n_{1}\sim n_{3}$)\par
$-0.9608\,\,\,-0.7721\,\,\,-0.7171$.\par
\hspace{-3mm}These solutions are found to be indistinguishable from those obtained using implicit or Crank-Nicholson time-integration schemes with $\Delta t=0.02$.\par

\newpage
\hspace{-3.6mm}\textbf{S-$9$ Properties and conditions of a four-node universe}\label{S-$8$ Properties and conditions of a four-node universe}\par

\begin{table}[h]
\footnotesize
\begin{center}
\begin{tabular}{lllll}
\hline
Node             &$1$ &$2$ &$3$ &$4$\\\hline
Material &Air   & Copper &Concrete &Helium \\\hline
$m$ ($kg$) &$10^{8}$   &$8.933$ &$0.0042$ &$10^{8}$ \\
$\rho$ ($kg/m^{3}$) &$1.160$   &$8933$ &$2100$ &$0.164$ \\
$c_{v}$ ($J/kgK$) &$717.5$   &$385.0$ &$1100$ &$3125.3$ \\
$\kappa$ ($W/mK$) &$0.026$   &$401$ &$0.810$ &$0.149$ \\  \hline
                  &$1/2\,\,\,\,\,\,\,\,0.2$   &$2/1\,\,\,\,\,\,0.2$   &$3/1\,\,\,\,\,\,0.1$ &$4/1\,\,\,\,\,\,0.01\,\,$ \\
$A$ ($m^{2}$)     &$1/3\,\,\,\,\,\,\,\,0.1$   &$2/3\,\,\,\,\,\,0.1$   &$3/2\,\,\,\,\,\,0.1$ &$4/2\,\,\,\,\,\,0.1\,\,$ \\
                  &$1/4\,\,\,\,\,\,\,\,0.01$   &$2/4\,\,\,\,\,\,0.1$   &$3/4\,\,\,\,\,\,0.2$ &$4/3\,\,\,\,\,\,0.2\,\,$ \\ \hline
                  &$1/2\,\,\,\,\,\,\,\,10$   &$2/1\,\,\,\,\,\,10$    &$3/1\,\,\,\,\,\,10$ &$4/1\,\,\,\,\,\,0\,\,$ \\
$h$ ($W/m^{2}K$)  &$1/3\,\,\,\,\,\,\,\,10$   &$2/3\,\,\,\,\,\,200$    &$3/2\,\,\,\,\,\,200$ &$4/2\,\,\,\,\,\,30\,\,$ \\
                  &$1/4\,\,\,\,\,\,\,\,0$   &$2/4\,\,\,\,\,\,30$    &$3/4\,\,\,\,\,\,30$ &$4/3\,\,\,\,\,\,30\,\,$ \\ \hline
$T_{o}$ ($K$)&$400$   &$320$ &$330$ &$300$ \\ \hline
\end{tabular}
\label{table3}
\end{center}
\end{table}

\end{document}